\documentclass[multphys,vecphys]{svmult}

\usepackage{makeidx}     
\usepackage{graphicx}    
\usepackage{multicol}    
\usepackage{amssymb}     

\makeindex             

\newcommand{\cd}{\cdot}

\newcommand{\OO}{\mathcal{O}}
\newcommand{\MM}{\mathcal{M}}

\newcommand{\HH}{\mathcal{H}}
\newcommand{\EE}{\mathcal{E}}
\newcommand{\al}{\alpha}
\renewcommand{\b}{\beta}
\newcommand{\de}{\delta}
\newcommand{\De}{\Delta}
\newcommand{\ep}{\epsilon}
\newcommand{\vep}{\varepsilon}
\newcommand{\ga}{\gamma}
\newcommand{\Ga}{\Gamma}
\newcommand{\ka}{\kappa}

\newcommand{\La}{\Lambda}
\newcommand{\la}{\lambda}
\newcommand{\Om}{\Omega}
\newcommand{\om}{\omega}
\newcommand{\si}{\sigma}
\newcommand{\Si}{\Sigma}
\renewcommand{\th}{\theta}
\newcommand{\vth}{\vartheta}
\newcommand{\vph}{\varphi}
\newcommand{\Up}{\Upsilon}
\newcommand{\ze}{\zeta}
\newcommand{\ra}{\rightarrow}

\newcommand{\lap}{\triangle}

\newcommand{\be}{\begin{equation}}
\newcommand{\ee}{\end{equation}}
\newcommand{\gsim}{\stackrel{>}{\sim}}
\newcommand{\lsim}{\stackrel{<}{\sim}}
\newcommand{\bea}{\begin{eqnarray}}
\newcommand{\eea}{\end{eqnarray}}
\newcommand{\bean}{\begin{eqnarray*}}
\newcommand{\eean}{\end{eqnarray*}}
\newcommand{\dd}{\partial}

\newcommand{\bk}{{\mathbf k}}
\newcommand{\bn}{{\mathbf n}}

\newcommand{\bx}{{\mathbf x}}

\newcommand{\ie}{{\em i.e. }}
\newcommand{\eg}{{\em e.g. }}

\newcommand{\da}{\dot{a}}
\newcommand{\lo}{{(\mathrm{long})}}
\newcommand{\mr}{\mathrm}
\newcommand{\dec}{\mathrm{dec}}
\def\nicefrac#1/#2{\leavevmode\kern.1em
\raise.5ex\hbox{\the\scriptfont0 #1}\kern-.1em
/\kern-.15em\lower.25ex\hbox{\the\scriptfont0 #2}}
\renewcommand{\l}{\left}
\renewcommand{\r}{\right}
\newcommand{\lan}{\langle}
\newcommand{\ran}{\rangle}

\begin{document}

\title*{Cosmological Perturbation Theory}%
\author{Ruth Durrer}
\institute{Universit\'e de Gen\`eve, D\'epartement de Physique Th\'eorique,\\
   24 Quai E. Ansermet, CH--1211 Gen\`eve, Switzerland \\
\texttt{ruth.durrer@physics.unige.ch}}



\maketitle

\begin{abstract}
This is a review on cosmological perturbation theory. After an
introduction, it presents the problem of gauge transformation. Gauge
invariant variables are introduced and the Einstein and conservation
equations are written in terms of these variables. Some examples,
especially perfect fluids and scalar fields are presented in detail. The
generation of perturbations during inflation is studied. Lightlike
geodesics and their relevance for CMB anisotropies are briefly
discussed. Perturbation theory in braneworlds is also introduced.
\end{abstract}

\begin{center}{\sl February 5, 2004} \end{center}

\newpage

\section{Introduction}
\index{introduction}
The idea that the large scale structure of our Universe might have
grown our of small initial fluctuations via gravitational instability
goes back to Newton (letter to Bentley, 1692\cite{NewtonB}).

The first relativistic treatment of linear perturbations in a
Friedmann-Lema\^\i tre universe was given by Lifshitz (1946)\cite{Lif46}.
There He found that  the gravitational potential cannot grow within
linear perturbation theory and he concluded that 
galaxies have not formed by gravitational instability.

Today we know that it is sufficient that matter density fluctuations can grow.
Nevertheless, considerable initial fluctuations with amplitudes of the order of
$10^{-5}$ are needed in order to reproduce the cosmic structures
observed today. These are much larger than typical statistical fluctuations
on scales of galaxies and we have to propose a mechanism to generate
them. Furthermore, the measurements
 of anisotropies in the cosmic microwave background show
that the amplitude of fluctuations is constant over a wide range of scales, 
the spectrum is scale independent.

As we shall see, standard inflation generically produces such a
spectrum of nearly scale invariant  fluctuations.

In this course I present gauge invariant cosmological perturbation
theory. I shall start by defining gauge invariant perturbation
variables. Then I present the basic perturbation equations. As
examples for the matter equations we shall consider perfect fluids and
scalar fields. The we briefly discuss lightlike geodesics and CMB
anisotropies (this section will be very brief since it is complemented
by the course on CMB anisotropies by A. Challinor). Finally, I shall
make some brief comments on perturbation theory for
braneworlds, a topic which is still wide open in my opinion.

\section{The background}
\index{background}
I shall not come back to the homogeneous universe which as been
discussed in depth in the course by K. Tamvakis.
 I just specify our notation which is as follows:

A Friedmann-Lema\^\i tre universe is a homogeneous and
isotropic solution of Einstein's equations. The hyper-surfaces
of constant time are homogeneous and isotropic, \ie, spaces
of constant curvature with metric
$a^2(\eta) \gamma_{ij} dx^i dx^j$, where $\ga_{ij}$ is the
metric of a space with constant curvature $\ka$. This metric can be
expressed in the form
\bea
\ga_{ij} dx^i dx^j &=& dr^2+\chi^2(r)\left(d\vth^2+sin^2\vth d\vph^2\right)\\
\chi(r) &=& \left\{ \begin{array}{lcl}
    r      & , & \ka=0 \\
    \sin r  & , & \ka=1 \\
    \sinh r & , & \ka=-1 , \end{array} \right.
\eea
where we have rescaled $a(\eta)$ such that $\ka=\pm1$ or $0$.
(With this normalization the scale factor  $a$ has the dimension of a
length and $\eta$ and $r$ are dimensionless for $\ka \neq 0$.) The
four-dimensional metric is then of the form
\be
g_{\mu\nu} dx^\mu dx^\nu = -a^2(\eta) d\eta^2
    +a^2(\eta) \ga_{ij} dx^i dx^j.
\ee
Here $\eta$ is called the {\em conformal time}.
The {\em physical} or {\em cosmological time} is given by  $dt =ad\eta$.

 Einstein's equations reduce to ordinary
differential equations for the function $a(\eta)$
(with $\dot{}\equiv d/d\eta$):
\bea
\left(\frac{\da}{a}\right)^2+\ka  = \HH^2 +\ka &=& \frac{8\pi G}{3} a^2 \rho
    + \frac{1}{3} \La a^2 \label{fl1} \\
\left(\frac{\da}{a}\right)^{\textstyle \cdot} = \dot\HH &=&
  -{4\pi G\over 3} a^2 \left(\rho+3 p\right)
    + \frac{1}{3} \La a^2 = \left(\frac{\ddot{a}}{a}\right)
    - \left(\frac{\da}{a}\right)^2 , \label{fl2}
\eea
where $\rho=-T^0_0$, $p=T^i_i$ (no sum!) and all other components of
the energy momentum tensor have to vanish by the requirement of
isotropy and homogeneity. $\La$ is the cosmological constant.
We have introduced $\HH=\dot a/a$. The Hubble parameter is defined by
\[  H={da/dt \over a} = {\dot a\over a^2} = \HH/a ~.\]

Energy momentum ``conservation'' (which is also a consequence of
(\ref{fl1})
and (\ref{fl2}) due to the contracted Bianchi identity) reads
\be
\dot{\rho}=-3\left(\frac{\da}{a}\right) (\rho+p) = -3(1+w)\HH , \label{emcons}
\ee
where $w\equiv p/\rho$.
Later we will also use $c_s^2\equiv \dot p/\dot\rho$. From the
definition of $w$ and $\rho$ together with Eq.~(\ref{emcons}) one
finds
\be
 \dot w = 3(1+w)(w-c_s^2)\HH ~.
\ee
From the Friedmann equations one easily concludes that for $\ka=\La=0$ and
$w=$const. the scale factor behaves like a power law,
\be \label{wconst}
a\propto \eta^{2\over 1+3w}\propto t^{2\over 3(1+w)} ~.
\ee
Important examples are
\be
a\propto \eta^q \quad \mbox{ with }\left\{ \begin{array}{lll}
q=2 & \mbox{for dust,} & w=0\\
q=1 & \mbox{for radiation,} & w=1/3\\
q=-1 & \mbox{for inflation (or a cosm. const.),} & w=-1\\
\end{array}\r.
\ee
We also define
\bea
\Om_\rho &=& \left(\frac{8\pi G\rho a^2}{3
  \HH^2}\right)_{\eta=\eta_0} \nonumber \\
\Om_\La &=& \left.\frac{\La a^2}
    {3 \HH^2}\right|_{\eta=\eta_0} \\
\Om_\ka &=& \left.\frac{-\ka}
    {\HH^2}\right|_{\eta=\eta_0} , \nonumber
\eea
where the index $_0$ indicates the value of a given variable today.
Friedmann's equation (\ref{fl1}) then requires
\be
1 = \Om_\rho + \Om_\La + \Om_\ka .
\ee
One often also uses $\Om = \Om_\rho+\Om_\La = 1-\Om_\ka$.

\section{Gauge invariant perturbation variables}
 \index{perturbation variables}
The observed Universe is not perfectly homogeneous and
isotropic. Matter is arranged in galaxies and clusters of galaxies and
there are large voids in the distribution of galaxies. Let us assume,
however, that these inhomogeneities lead only to small variations of
the geometry which we shall treat in first order perturbation
theory. For this we define the perturbed ¨geometry by 
\be
g_{\mu\nu} = \bar{g}_{\mu\nu}+\vep a^2 h_{\mu\nu}
\ee
$\bar{g}_{\mu\nu}$ being the unperturbed Friedmann metric.
We conventionally set (absorbing the ``smallness''
parameter $\vep$ into $h_{\mu\nu}$)
\be \begin{array}{llll}
g_{\mu\nu} = \bar{g}_{\mu\nu} + a^2 h_{\mu\nu}, &
    \quad \bar{g}_{00}=-a^2, &
    \quad \bar{g}_{ij}=a^2 \ga_{ij}&
    \quad |h_{\mu\nu}| \ll 1 \\[.2cm]
T^\mu_\nu = \overline{T}^\mu_\nu + \th^\mu_\nu,&
    \quad \overline{T}^0_0 = -\bar{\rho},&
    \quad \overline{T}^i_j = \bar{p} \de^i_j&
    \quad |\th^\mu_\nu|/\bar{\rho} \ll 1 .
\end{array} \ee

\subsection{Gauge transformation, gauge invariance}

The first fundamental problem we want to discuss is the 
choice of gauge in cosmological perturbation theory:

For linear perturbation theory to apply, the spacetime manifold
$\MM$ with metric $g$ and the energy momentum tensor $T$ of the real,
observable universe must be in some sense close to a Friedmann universe,
\ie,
the manifold $\MM$ with a Robertson--Walker metric $\bar{g}$ and a
homogeneous and isotropic energy momentum tensor $\overline{T}$. It
is an interesting, non--trivial unsolved problem how to construct 'the
best' $\bar{g}$ and $\overline{T}$ from the physical fields $g$ and
$T$ in practice.  There are two main
difficulties: First, spatial averaging procedures depend on the choice of a
hyper--surface of constant time
 and they do not commute with derivatives, so that averaged
 fields $\bar{g}$ and $\overline{T}$
 will in general not satisfy Einstein's equations. Secondly, averaging
is in practice impossible over super--horizon scales.

Even though we cannot give a constructive prescription,
we now assume that there exists an averaging procedure which
 leads to a Friedmann
universe with spatially averaged tensor fields $\overline{Q}$, such that
the deviations
$(T_{\mu\nu}-\overline{T}_{\mu\nu})/\max_{\{\al\b\}}\{|\overline{T}_{\al\b}|\}$
and
$(g_{\mu\nu}-\overline{g}_{\mu\nu})/\max_{\{\al\b\}}\{\overline{g}_{\al\b}\}$
  are small, and $\bar{g}$ and $\overline{T}$ satisfy Friedmann's equations.
Let us call such an averaging procedure 'admissible'.
There may be many different admissible averaging procedures (e.g. over
different hyper--surfaces) leading to slightly different Friedmann backgrounds.
But since $|g-\bar{g}|$ is small of order $\ep$, the difference of the
two Friedmann backgrounds must also be small of order  $\ep$ and we
can regard it as part of the perturbation.

We consider now a fixed
admissible Friedmann background $(\bar{g}, \bar{T})$ as chosen. Since
the theory is invariant under diffeomorphisms (coordinate
transformations),  the perturbations are not unique. For an arbitrary
diffeomorphism $\phi$ and its push forward $\phi^*$, the two metrics
$g$ and $\phi^*(g)$ describe the same geometry. Since we have chosen
the background metric $\bar{g}$ we only allow diffeomorphisms which
leave  $\bar{g}$ invariant {\em i.e.} which deviate only in first
order form the identity. Such an 'infinitesimal' diffeomorphism
can be represented as the infinitesimal flow of a vector field $X$,
$\phi = \phi_\ep^X$. Remember the definition of the flow: For the integral
curve $\ga_x(s)$ of $X$ with starting point $x$, i.e., $\ga_x(s=0)=x$
we have $\phi_s^X(x) = \ga_x(s)$. In terms of the vector field $X$, to
 first order in $\ep$, its pullback is then
of the form
\[ \phi^* = id +\ep L_X \]
($L_X$ denotes the Lie derivative in
direction $X$). The transformation $g\ra \phi^*(g)$  is
equivalent to $\bar g +\ep a^2 h \ra \bar g +\ep(a^2h+L_X\bar g) + \OO(\ep^2)$,
{\em i.e.} under an 'infinitesimal coordinate transformation' the
metric perturbation $h$ transforms as
\be
 h \ra h + a^{-2}L_X\bar{g} ~. \label{2gt}
\ee
In the context of cosmological perturbation theory, infinitesimal
coordinate transformations are called 'gauge transformations'. The
perturbation of a arbitrary tensor field $Q=\bar Q +\ep Q^{(1)}$ obeys
the gauge transformation law
\be
 Q^{(1)}  \ra  Q^{(1)} + L_X\bar{Q}~.  \label{gaugeQ}
\ee

Since every vector field $X$ generates a gauge
transformation $\phi = \phi_\ep^X$, we can conclude that only perturbations
 of tensor fields with  $ L_X\overline{Q}=0$ for all vector fields $X$,
i.e., with vanishing (or constant) 'background contribution' are gauge
invariant. This simple result is sometimes referred to as the
'Stewart-Walker Lemma' \cite{StW}.

The gauge dependence of perturbations has caused many controversies in the
literature, since it is often difficult to extract the physical
meaning of gauge dependent perturbations, especially on super--horizon
scales. This has led to the development of gauge invariant
perturbation theory
which we are going to use throughout this review. The advantage of the
gauge--invariant formalism is that the variables used have simple geometric
and physical meanings and are not plagued by gauge modes.
Although the derivation requires somewhat more work, the final system
of perturbation equations is usually simple and well suited for numerical
treatment. We shall also see, that on sub-horizon scales, the
gauge invariant matter perturbation variables approach the usual, gauge
dependent ones. Since one of the gauge invariant geometrical perturbation
variables corresponds to the Newtonian potential, the Newtonian limit can be
performed easily.

First we note that since all  relativistic
equations are covariant (i.e. can be written in the form $Q=0$ for some
tensor field $Q$), it is always possible to express the corresponding
perturbation equations in terms of gauge invariant variables
\cite{Ba,KS,Rfund}.

\subsection{Harmonic decomposition of perturbation variables}

Since the $\{\eta=\mr{const}\}$ hyper-surfaces are homogeneous and
isotropic, it is sensible to perform a harmonic analysis: A
(spatial) tensor field $Q$ on these hyper-surfaces can be decomposed
into components which transform irreducibly under translations and
rotations. All such components evolve independently. For a scalar
quantity $f$ in the case $\ka=0$ this is nothing else than its
Fourier decomposition:
\be
f(\bx,\eta)=\int d^3\!k \hat{f}(\bk) e^{i\bk\bx}.
\ee
(The exponentials $Y_\bk(\bx)=e^{i\bk\bx}$ are the unitary irreducible
representations of the Euclidean translation group.) For $\ka=1$ such a
decomposition also exists, but the values $k$ are discrete,
$k^2=\ell(\ell+2)$ and for $\ka=-1$, they are bounded from below,
$k^2>1$. Of course, the functions $Y_\bk$ are different for $\ka\neq 0$.

They form the complete orthogonal set of eigenfunctions of the
Laplacian,
\be
\Delta Y_\bk^{(S)}=-k^2 Y_\bk^{(S)} .
\ee

In addition, a tensorial variable (at fixed position $\bx$) can be
decomposed into irreducible components under the rotation
group $SO(3)$.

For a vector field, this is its decomposition
into a gradient and a rotation,
\be
V_i = \nabla_i \vph + B_i ,\label{vdec} 
\ee
where
\be
B^i_{|i}=0 ,
\ee
where we used $X_{|i}$ to denote the three--dimensional covariant
derivative of $X$. Here
$\vph$ is the spin 0 and ${\bf B}$ is the spin 1 component
of the vector field V.

For a symmetric tensor field we have
\be
H_{ij} = H_L \ga_{ij} +\left(\nabla_i\nabla_j-\frac{1}{3}\De\ga_{ij}\right)H_T
    +\frac{1}{2}\left(H^{(V)}_{i|j}+H^{(V)}_{j|i}\right) + H^{(T)}_{ij},
\label{tdec}
\ee
where
\be
H^{(V)|i}_i=H^{(T)^i}_i=H^{(T)^j}_{i|j}=0.
\ee
Here $H_L$ and $H_T$ are spin 0 components, $H^{(V)}_i$ is a
spin 1 component and $H^{(T)}_{ij}$ is a spin 2 component.

We shall not need higher tensors (or spinors).
As a basis for vector and tensor modes we use the vector and tensor
type eigenfunctions of the La\-placian,
\bea
 \Delta Y_j^{(V)} &=&-k^2 Y_j^{(V)} \quad \mbox{ and} \\
 \Delta Y_{ji}^{(T)} &=&-k^2 Y_{ji}^{(T)}~,
\eea
where $Y_j^{(V)}$ is a transverse vector,  $Y_j^{(V)|j}=0$ and $
Y_{ji}^{(T)} $ is a symmetric transverse traceless tensor, $Y_j^{(T)j}=
Y_{ji}^{(T)|i}= 0$.

According to Eqs.~(\ref{vdec}) and (\ref{tdec}) we can construct scalar type
vectors and tensors and vector type tensors. To this goal we define
\bea
Y^{(S)}_j &\equiv& -k^{-1}Y^{(S)}_{|j} \\
Y^{(S)}_{ij} &\equiv& k^{-2}Y^{(S)}_{|ij} +{1\over 3}\ga_{ij}Y^{(S)} \\
Y^{(V)}_{ij} &\equiv& -{1\over 2k}(Y^{(V)}_{i|j} +Y^{(V)}_{j|i}) ~.
\eea
In the following we shall extensively use this decomposition
and write down the perturbation equations for a given mode
$k$.

The decomposition of a vector field is then of the form
\be
B_i = B Y^{(S)}_i+B^{(V)} Y^{(V)}_i. \label{vdec2}
\ee
The decomposition of a tensor field is given by (compare Eq.~(\ref{tdec}))
\be
H_{ij}=H_LY^{(S)} \ga_{ij} +H_TY^{(S)}_{ij}
    +H^{(V)}Y^{(V)}_{ij} + H^{(T)}Y^{(T)}_{ij}.
\label{tdec2}
\ee
Here $B$, $B^{(V)}$, $H_L$, $H_T$, $H^{(V)}$ and $H^{(T)}$ are
functions of $\eta$ and $\bk$

\subsection{Metric perturbations}
Perturbations of the metric are of the form
\be
g_{\mu\nu} = \bar{g}_{\mu\nu} + a^2 h_{\mu\nu} .
\ee
We parameterize them as
\be
h_{\mu\nu} dx^\mu dx^\nu = -2 A d\eta^2 - 2B_i d\eta dx^i
    +2 H_{ij} dx^i dx^j, \label{defpert}
\ee
and we decompose the perturbation variables $B_i$ and $H_{ij}$ according  to
(\ref{vdec2}) and (\ref{tdec2}).

Let us consider the behavior of $h_{\mu\nu}$ under gauge
transformations. We set the vector field defining the gauge
transformation to
\be
{\bf X} = T\dd_\eta+L^i\dd_i .
\ee
Using simple identities from differential geometry like
$L_{\bf X}(df)=d(L_{\bf X} f)$ and \\
$(L_{\bf X}\ga)_{ij}=X_{i|j}+X_{j|i}$, we obtain
\bea
L_{\bf X} \bar{g} &=& a^2\left[-2\left(\frac{\da}{a}T+\dot{T}\right)d\eta^2
    +2\left(\dot{L}_i-T_{,i}\right)d\eta dx^i \right. \nonumber \\ &&
 \left.    +\left(2\frac{\da}{a}T\ga_{ij}+L_{i|j}+L_{j|i}\right)dx^i dx^j\right].
\eea

Comparing this with (\ref{defpert}) and using (\ref{2gt}) we
obtain the following behavior of our perturbation variables under
gauge transformations (we decompose the vector 
$L_i=LY^{(S)}_i+L^{(V)}Y^{(V)}_i$):
\bea
A &\rightarrow& A + \frac{\da}{a}T + \dot{T} \label{transA}\\
B &\rightarrow& B - \dot{L} -k T\\
B^{(V)} &\rightarrow& B^{(V)} - \dot{L}^{(V)}\\
H_L &\rightarrow& H_L + \frac{\da}{a}T + \frac{k}{3} L\\
H_T &\rightarrow& H_T - kL \label{transHT}\\
H^{(V)} &\rightarrow& H^{(V)} -k L^{(V)} \\
H^{(T)} &\rightarrow& H^{(T)} .
\eea
Two scalar and one vector variable can be set to zero
by a cleverly chosen gauge transformations.

One often chooses $kL=H_T$ and $kT=B+\dot{L}$, so that the variables
$H_T$ and $B$ vanish. In this gauge (longitudinal gauge), scalar
perturbations of the metric are of the form ($H_T=B=0$):
\be
h_{\mu\nu}^{(S)} = -2 \Psi d\eta^2 - 2 \Phi \ga_{ij} dx^i dx^j .
\ee

$\Psi$ and $\Phi$ are the so called {\em Bardeen} potentials.
In a generic gauge  they are given by
\bea
\Psi &=& A - \frac{\da}{a}k^{-1} \sigma -k^{-1}\dot{\sigma} \\
\Phi &=& - H_L - \frac{1}{3} H_T +\frac{\da}{a}k^{-1} \sigma
\eea
with $\si=k^{-1}\dot{H}_T-B$. A short calculation using Eqs. (\ref{transA})
to (\ref{transHT}) shows that $\Psi$ and $\Phi$  are gauge invariant.

In a Friedmann universe the Weyl tensor vanishes. It therefore is a
gauge invariant perturbation. For scalar perturbations one finds
\be
C^0_{~ i0j} = {1\over 2}\left[ (\Psi+\Phi)_{|ij}
-{1\over 3}\lap(\Psi+\Phi)\ga_{ij}\right] \ee
All other components vanish.

For vector perturbations it is convenient to set $kL^{(V)}=H^{(V)}$
so that $H^{(V)}$ vanishes and we have
\be
h_{\mu\nu}^{(V)} dx^\mu dx^\nu = 2 \si^{(V)}Y^{(V)}_i d\eta dx^i.
 \label{vecg}\ee
We shall call this gauge the ``vector gauge''.
In general $\si^{(V)}=k^{-1}\dot{H}^{(V)}-B^{(V)}$ is gauge invariant\footnote{
$Y^{(V)}_{ij}\si^{(V)}$ is the shear of the hyper-surfaces
of constant time.}.
The Weyl tensor from vector perturbation is given by
\begin{eqnarray}
C^0_{~ i0j} &=& {1\over 2}\dot\si^{(V)}Y^{(V)}_{ij}\\
C^0_{~ jlm} &=& {1\over 2}\si^{(V)}[Y^{(V)}_{jl|m}- Y^{(V)}_{jm|l}
  -{1\over 3}\ga_{jl}Y^{(V)}_{m|k}{}^k + {1\over 3}\ga_{jm}Y^{(V)}_{l|k}{}^k  ]
\end{eqnarray}
Clearly there are no tensorial (spin 2) gauge transformation
and hence $H_{ij}^{(T)}$ is gauge invariant.
The expression for the Weyl tensor from tensor perturbation is identical
to the one for vector perturbation upon replacement of $\si^{(V)}_{ij}$
by $\dot H_{ij}^{(T)}$.

\subsection{Perturbations of the energy momentum tensor}

Let $T^\mu_\nu=\overline{T}^\mu_\nu+\th^\mu_\nu$ be the full
energy momentum tensor. We define its energy density $\rho$
and its energy flow 4-vector $u$ as the time-like eigenvalue
and eigenvector of $T^\mu_\nu$:
\be
T^\mu_\nu u^\nu = -\rho u^\mu , \quad u^2=-1.
\ee

We then parameterize their perturbations by
\be
\rho=\bar{\rho}\left(1+\de\right), \quad u=u^0 \dd_t + u^i \dd_i .
\ee
$u^0$ is fixed by the normalization condition,
\be
u^0 = \frac{1}{a} (1-A) .
\ee
We further set
\be
u^i = \frac{1}{a} v^i = vY^{(S)i} + v^{(V)}Y^{(V)i}.
\ee

We define $P^\mu_\nu\equiv u^\mu u_\nu+\de^\mu_\nu$, the
projection tensor onto the part of tangent space normal to
$u$ and set the stress tensor
\be
\tau^{\mu\nu}=P^\mu_\al P^\nu_\b T^{\al\b} .
\ee

In the unperturbed case we have
$\tau^0_0=0, \tau^i_j=\bar{p} \de^i_j$. Including
 perturbations, to first order we still obtain
\be
\tau^0_0=\tau^0_i=\tau^i_0=0.
\ee

But $\tau^i_j$ contains in general perturbations. We set
\be
\tau^i_j = \bar{p}\left[\left(1+\pi_L\right)\de^i_j+\Pi^i_j\right]
    , \quad \mr{with} \quad \Pi^i_i=0.
\ee
We decompose $\Pi^i_j$ as
\be
\Pi^i_j =  \Pi^{(S)}Y^{(S)\,i}_j
    + \Pi^{(V)}Y^{(V)\,i}_j
    + \Pi^{(T)} Y^{(T)\,i}_j.
\ee

We shall not derive the gauge transformation properties of these
perturbation variables in detail, but just state some results which
can be obtained as an exercise (see also~\cite{KS}):
\begin{itemize}
\item Of the variables defined above only the $\Pi^{(S,V,T)}$
are gauge invariant; they describe the anisotropic stress tensor,
$\Pi^\mu_\nu=\tau^\mu_\nu-\nicefrac1/3\tau^\al_\al\de^\mu_\nu$.
They are gauge invariant due to the Stewart--Walker
lemma, since $\bar{\Pi}=0$. For perfect fluids $\Pi^\mu_\nu=0$.
\item A second gauge invariant variable is
\be
\Ga=\pi_L-\frac{c_s^2}{w}\de,
\ee
where $c_s^2\equiv\dot{p}/\dot{\rho}$ is the adiabatic sound speed
and $w\equiv p/\rho$ is the enthalpy. One can show that $\Ga$ is
proportional to the divergence of the entropy flux of the
perturbations. Adiabatic perturbations are characterized
by $\Ga=0$.
\item Gauge invariant density and velocity perturbations
can be found by combining $\de$, $v$ and $v_i^{(V)}$ with
metric perturbations.
\end{itemize}

We shall use
\bea
V &\equiv& v - \frac{1}{k} \dot{H}_T = v^\lo \\[2mm]
D_s &\equiv& \de + 3(1+w){\dot a\over a}(k^{-2}\dot{H}_T -
k^{-1}B) \equiv \de^\lo \\
 D &\equiv & \de^\lo+3(1+w)
\left(\frac{\da}{a}\right)\frac{V}{k}
\label{Ddef}\\
D_g &\equiv& \de + 3(1+w) \left(H_L+\frac{1}{3}
H_T\right)=\de^\lo - 3(1+w)\Phi  \label{Dgdef}\\[2mm]
 V^{(V)} &\equiv& v^{(V)} - \frac{1}{k}\dot{H}^{(V)}
    = v^{(\mr{vec})}\\
\Om &\equiv& v^{(V)} - B^{(V)} =  v^{(\mr{vec})}-B^{(V)}\\
 \Om - V^{(V)} &=&\si^{(V)} .
\eea
Here $ v^\lo,~\de^\lo$ and $ v_i^{(\mr{vec})}$ are the velocity (and
density) perturbations in the longitudinal and vector gauge
respectively, and $\si^{(V)}$ is the metric perturbation in vector
gauge (see Eq.~(\ref{vecg})).
 These variables
 can be  interpreted nicely in terms of gradients
of the energy density and the shear and vorticity of the velocity
field~\cite{Ellis}.

Here I just want to show that on scales much smaller than the
Hubble scale, $k\eta\gg 1$, the metric perturbations are much
smaller than $\de$ and $v$ and we can thus ``forget them'' (which
will be important when comparing experimental results with
calculations in this formalism):

The perturbations of the Einstein tensor are given by
second derivatives of the metric perturbations. Einstein's
equations yield the following order of magnitude estimate:
\bea
\OO\left(\frac{\de T}{T}\right)
\underbrace{\OO\left(8\pi GT\right)}_{\OO\left(\frac{\da}{a}\right)^2=\OO\left(\eta^{-2}\right)}
&=& \OO\left(\frac{1}{\eta^2}h+\frac{k}{\eta}h+k^2 h\right)\\
\OO\left(\frac{\de T}{T}\right) &=& \OO\left(h+k\eta h+(k\eta)^2 h\right) .
\eea
For $k\eta\gg 1$ this gives $\OO(\de,v)=\OO\left(\frac{\de T}{T}\right)\gg\OO(h)$.
On sub-horizon scales the difference between $\de$, $\de^\lo$, $D_g$
and $D$ is negligible as well as the difference between $v$ and $V$
or $v^{(V)}$, $V^{(V)}$ and $\Om^{(V)}$.

\section{Einstein's equations}
\index{Einstein equations}
We do not derive the first order perturbations of Einstein's equations. This can be
done by different methods, for example with Mathematica. We just
write down those equations which we shall need later.

\subsection{Constraint equations}
\bea \left. \begin{array}{rclc}
4\pi G a^2 \rho D & = & -(k^2 -3\ka)\Phi & (00)\\
4\pi G a^2 (\rho+p) V & = & k\left(\left(\frac{\da}{a}\right)\Psi+
\dot{\Phi}\right) & (0i) 
\end{array} \right\} && \mr{(scalar)}\label{scalcntr} \\
8\pi G a^2 (\rho+p)\Om = \frac{1}{2} \left(2\ka-k^2\right)\si^{(V)} \quad (0i)
\quad && \mr{(vector)} \label{veccntr}
\eea

\subsection{Dynamical equations}
\bea
k^2 \left(\Phi-\Psi\right) &=& 8\pi G a^2 p\Pi^{(S)}
 \quad\quad \mr{(scalar)} \label{scaldyn}\\
k\left(\dot{\si}^{(V)}+2\left(\frac{\da}{a}\right)\si^{(V)}\right) &=&
8\pi G a^2 p \Pi^{(V)}
    \quad\quad\mr{(vector)} \label{vecdyn} \\
\ddot{H}^{(T)}
    +2\left(\frac{\da}{a}\right)\dot{H}^{(T)}
    +\left(2\ka+k^2\right)H^{(T)}  &=&
8\pi G a^2 p \Pi_{ij}^{(T)}
\quad\quad \mr{(tensor)} \label{tensdyn}
\eea
There is a second dynamical scalar equation, which is somewhat combersome and
not really needed, since we may use one of the conservation eqs. given
below instead. Note that for perfect fluids, where $\Pi^i_j\equiv0$, we have
$\Phi=\Psi$, $\si^{(V)}\propto 1/a^2$ and $H^{(T)}$ obeys a damped
wave equation. The damping term can be neglected on small scales (over
short time periods) when
$\eta^{-2}\lsim 2\ka+k^2$, and $H_{ij}$ represents propagating
gravitational waves. For vanishing curvature, these are just the
sub-horizon scales, $k\eta\gsim 1$. For $\ka<0$, waves oscillate with
a somewhat smaller frequency, $\om=\sqrt{2\ka+k^2}$,
 while for $\ka>0$ the frequency is somewhat larger than $k$.

\subsection{Energy momentum conservation}
The conservation equations, $T^{\mu\nu}_{;\nu}=0$ lead to the following
perturbation  equations.
\bea \hspace{-1cm}\left. \begin{array}{c}
\dot{D}_g+3\left(c_s^2-w\right)\left(\frac{\da}{a}\right)D_g
+(1+w)kV+3w\left(\frac{\da}{a}\right)\Ga=0 \\
\dot{V}+\left(\frac{\da}{a}\right)\left(1-3c_s^2\right)V = k\left(\Psi
+ 3c_s^2\Phi\right) + \frac{c_s^2 k}{1+w}D_g \\
 \hspace{2cm} +\frac{wk}{1+w}\left[\Ga-\frac{2}{3}\left(1 -
 \frac{3\ka}{k^2}\right)\Pi\right] 
\end{array} \right\} && \mr{(scalar)} \label{scalcons}\\
\hspace{-1cm}\dot{\Om}_i+\left(1-3c_s^2\right)\left(\frac{\da}{a}\right)\Om_i
=\frac{p}{2(\rho+p)}\left(k-\frac{2\ka}{k}\right)\Pi_i^{(V)} \qquad &&
\mr{(vector)} \label{veccons}
\eea
These can of course also be obtained from the Einstein equations since they are
equivalent to the contracted Bianchi identities. For scalar
perturbations we have 4 independent equations and 6
variables. For vector perturbations we have 2 equations and 3
variables, while for tensor perturbations we have 1 equations and 2
variables. To close the system we must add some matter equations. The
simplest prescription is to  set $\Ga=\Pi_{ij}=0$. This matter
equation, which describes adiabatic perturbations of a perfect fluid gives us
exactly two additional equations for scalar perturbations and one each
for vector and tensor perturbations.

Another simple example is a universe with matter content given by a
scalar field. We shall discuss this case in the next section. More
complicated examples are those of several interacting particle species of
which some have to be described by a Boltzmann equation. This is the
actual universe at late times, say $z\lsim 10^7$.

\subsection{A special case}
Here we want to rewrite the scalar perturbation equations for a simple but
important special case.  We consider adiabatic perturbations of
a perfect fluid. In this case $\Pi=0$ since there are no anisotropic
stresses and $\Ga=0$. Eq.~(\ref{scaldyn}) then implies $\Phi=\Psi$. Using
the first equation of (\ref{scalcntr}) and
Eqs. (\ref{Dgdef},\ref{Ddef}) to replace $D_g$ in the second of
Eqs.~(\ref{scalcons}) by $\Psi$ and $V$, finally
replacing $V$  by (\ref{scalcntr}) one can derive a second order
equation for $\Psi$, which is, in this case the only dynamical degree
of freedom
\be \label{Psifluid}
\ddot\Psi + 3\HH(1+c_s^2)\dot\Psi +[(1+3c_s^2)(\HH^2-\ka) -(1+3w)\HH^2
  +c_s^2k^2]\Psi = 0~.
\ee

Another interesting case (especially when discussing inflation) is the
scalar field case. There, as we shall see in Section~\ref{scalarfield}, $\Pi=0$, but in
general $\Ga\neq 0$ since $\de p/\de\rho\neq \dot
p/\dot\rho$. Nevertheless, since this case again has only one
dynamical degree of freedom, we can express the perturbation equations
in terms of one single second order equation for $\Psi$. In
Section~\ref{scalarfield} 
we shall find the following equation for a perturbed scalar field cosmology
\be \label{Psifield}
\ddot\Psi + 3\HH(1+c_s^2)\dot\Psi +[(1+3c_s^2)(\HH^2-\ka) -(1+3w)\HH^2
  +k^2]\Psi = 0~.
\ee

The only difference between the perfect fluid and scalar field
perturbation equation is that the latter is missing the factor $c_s^2$
in front of the oscillatory $k^2$ term. Note also that for $\ka=0$ and
$w=c_s^2=$ constant the time dependent mass term
$m^2(\eta)=-(1+3c_s^2)(\HH^2-\ka) + (1+3w)\HH^2$ vanishes.
It is useful to define also the variable~\cite{Mukhanov:1992tc}
\be\label{defu}
 u = a\left[4\pi G(\HH^2-\dot\HH+\ka)\right]^{-1/2}\Psi,  \ee
which satisfies the equation
\be \label{ueq}
\ddot u +(\Upsilon k^2 -\ddot\th/\th)u = 0,
\ee
where $\Up = c_s^2$ or $\Up = 1$ for a perfect fluid or a scalar field
background respectively, and
\be
 \th = {3\HH\over 2a\sqrt{\HH^2-\dot\HH+\ka}}~.
\ee

Another interesting variable is
\be\label{curva}
\ze \equiv {2(\HH^{-1}\dot\Psi +\Psi)\over 3(1+w)} +\Psi ~.
\ee
Using Eqs.~(\ref{Psifluid}) and (\ref{Psifield}) respectively one obtains
\be \label{cons}
\dot\ze = -k^2{\Up\HH\over \HH^2-\dot\HH}\Psi ~,
\ee
hence on super horizon scales, $k/\HH\ll 1$, this variable is
conserved.

The evolution of $\ze$ is closely related to the canonical variable
$v$ defined by
\be \label{defv}
v = -{a\sqrt{\HH^2-\dot\HH}\over \sqrt{4\pi G}\Up\HH}\ze~.
\ee
which satisfies the equation
\be
 \ddot v +(\Up k^2- \ddot z/z)v =0~,  \label{veq1}
\ee
for
\be  \label{z}
z = {a\sqrt{\HH^2-\dot \HH +\ka}\over \Up \HH}~.
\ee
More details on the significance of the canonical variable $v$ will be
found in sections~\ref{scalarfield} and~\ref{sec:inf}.

\section{Simple examples}
We first discuss two simple applications which
are important to understand the CMB anisotropy spectrum. 

\subsection{The pure dust fluid for $\ka=0, \La=0$}
We assume the dust to have $w=c_s^2=p=0$ and $\Pi=\Ga=0$.
Equation (\ref{Psifluid}) then reduces to 
\be \label{Psidust}
\ddot\Psi + {6\over \eta}\dot\Psi  = 0~,
\ee
with the general solution
\be
\label{Psidust2}
\Psi = \Psi_0 +\Psi_1{1\over \eta^5} 
\ee
with arbitrary constants $\Psi_0$ and $\Psi_1$.
Since the perturbations are supposed to be small initially,
they cannot diverge for $\eta \ra 0$, and we have therefore
to choose the growing mode, $\Psi_1=0$. Another way to argue is as
follows: If the 
mode $\Psi_1$ has to be small already at some early initial time
$\eta_\mr{in}$, it will be even much smaller later and may
hence be neglected at late 
times. But also the  $\Psi_0$ mode is only constant. This fact led
Lifshitz who was the first to analyze cosmological perturbations to
the conclusions that linear perturbations do not grow in a Friedman
universe and cosmic structure cannot have evolved by gravitational
instability~\cite{Lif46}. However, the important point to note here
is that, even if the gravitational potential remains constant, matter density
fluctuations do grow on sub-horizon scales and therefore
inhomogeneities, structure can evolve on scales which are smaller
than the Hubble scale. To see that we consider the conservation equations
(\ref{scalcons}), (\ref{scaldyn}) and the Poisson equation
(\ref{scalcntr}). For the pure dust case, $w=c_s^2=\Pi=\Ga=0$, they reduce to
\bea
\dot{D}_g&=&-kV\quad\mr{(energy~conservation)} \label{dust1}\\
\dot{V}+\l(\frac{\da}{a}\r)V &=& k\Psi \quad
	\mr{(gravitational~acceleration)}  \label{dust2}\\
-k^2\Psi &=&4\pi
 Ga^2\rho\l(D_g+3\l(\Psi+\l(\frac{\da}{a}\r)k^{-1}V\r)\r)
   \quad\mr{(Poisson)} , \label{dust4}
\eea
where we have used the relation
\be
D=D_g+3(1+w)\l(-\Phi+\l(\frac{\da}{a}\r)k^{-1}V\r) . \label{DDgrel}
\ee
The Friedmann equation for dust gives
 $4\pi G\rho a^2=\nicefrac3/2(\da/a)^2=6/\eta^2$.
Setting $k\eta=x$ and $'=d/dx$, the system (\ref{dust1}-\ref{dust4})
becomes
\bea
D'_g&=&-V\label{dustb1} \\
V'+\frac{2}{x}V&=&\Psi\label{dustb2}\\
\frac{6}{x^2}\l(D_g+3\l(\Psi+\frac{2}{x}V\r)\r)&=& -\Psi. \label{dustb3}
\eea

We use (\ref{dustb3}) to eliminate $\Psi$ and (\ref{dustb1})
to eliminate $D_g$, leading to
\be
\l(18+x^2\r)V''+\l(\frac{72}{x}+4 x\r)V'-\l(\frac{72}{x^2}+4\r)V=0.
\label{Vpp}\ee
The general solution of Eq.~(\ref{Vpp}) is
\be
V=V_0 x+\frac{V_1}{x^4}
\ee
The $V_1$ mode is the decaying mode (corresponding to $\Psi_1$) which
we neglect. 
The perturbation variables are then given by
\bea
V &=& V_0 x \\
D_g &=& -15 V_0 -\frac{1}{2}V_0 x^2\\
V_0 &=& \Psi_0/3 .
\eea

We distinguish two regimes:\\
{\bf \em i}) super-horizon, $x\ll1$ where we have
\bea
V&=&{1\over 3}\Psi_0x\\
D_g&=&-5\Psi_0\\
\Psi&=&\Psi_0~.
\eea
Note that even though $V$ is growing, it always remains much small
than $\Psi$ or $D_g$ on super-horizon scales. Hence the largest
fluctuations are of order $\Psi$ which is constant.\\
{\bf \em ii}) Sub-horizon, $x\gg1$ where the solution is  dominated by 
the terms
\bea
V&=&{1\over 3}\Psi_0x\\
D_g&=&-\frac{1}{6}\Psi_0 x^2\\
\Psi&=&\Psi_0=\mr{constant.} \label{eq48}
\eea

Note that for dust
\[ D =D_g + 3\Psi +{6\over x}V = -\frac{1}{6}\Psi_0 x^2 ~. \]
In the variable $D$ the constant term has disappeared and we have $D\ll \Psi$ 
on super-horizon scales, $x\ll 1$.

On sub-horizon scales, the density fluctuations grow like the scale
factor $\propto x^2\propto a$. Nevertheless, Lifshitz'
conclusion~\cite{Lif46} that pure gravitational
instability cannot be the cause for structure formation has some truth:
If we start from tiny thermal fluctuations of the order
of $10^{-35}$, they can only grow to about
$10^{-30}$ due to this mild, power law instability during the matter dominated
regime. Or, to put it differently,
if we want to form structure by gravitational instability,
we need initial fluctuations of the order of at least $10^{-5}$, much
larger than thermal fluctuations.
One possibility to create such fluctuations is quantum particle
production in the classical gravitational field during inflation.
 The rapid expansion of the universe during inflation quickly expands 
microscopic scales at which quantum fluctuations are important to 
cosmological scales where these fluctuations are then ``frozen in'' as 
classical perturbations in the energy density and the geometry.
We will discuss the induced spectrum on fluctuations in Section~\ref{sec:inf}.

\subsection{The pure radiation fluid, $\ka=0, \La=0$}

In this limit we set $w=c_s^2=\nicefrac1/3$ and $\Pi=\Ga=0$ so that
$\Phi=-\Psi$. We conclude from $\rho\propto a^{-4}$ that $a\propto\eta$.
For radiation, the $u$--equation (\ref{ueq}) becomes
\be \label{ueq2}
\ddot u +({1\over 3} k^2 - { 2\over \eta^2})u = 0, 
\ee
with general solution
\be \label{urad}
 u(x) = A\l({\sin(x)\over x} -\cos(x)\r) + B\l({\cos(x)\over x}
 -\sin(x)\r) ~,
\ee
where we have set $x=k\eta/\sqrt{3}=c_sk\eta$. For the Bardeen
potential we obtain with (\ref{defu}), up to constant factors,
\be \label{Psirad}
 \Psi(x) ={u(x) \over x^2} ~.
\ee
On super-horizon scales, $x\ll 1$, we have
\be
 \Psi(x) \simeq {A\over 3} + {B \over x^3} ~.
\ee
We assume that the perturbations have been initialized at some early
time $x_\mr{in} \ll 1$ and that at this time the two modes have been
comparable. If this is the case then $B\ll A$ and we may neglect the
$B$-mode at later times.

To determine the density and velocity perturbations and for
illustration, we also solve the radiation equations using the
conservation and Poisson equations like for dust. In the radiation case
the perturbation
equations become (with the same notation as above, $x=c_sk\eta$)
\bea
D'_g&=&-\frac{4}{\sqrt{3}} V\\
V' &=& 2\sqrt{3}\Psi+\frac{\sqrt{3}}{4}D_g\\
-2x^2\Psi &=& D_g+4\Psi+\frac{4}{\sqrt{3}x}V ~ . \label{radflu3}
\eea
The general solution of this system is
\bea
D_g &=& D_2 \l[\cos(x) 	-{2\over x}\sin(x)\r]
	\nonumber\\ &&
+ D_1 \l[\sin(x) +{2\over x}\cos(x)\r]\\
V&=&-\frac{\sqrt{3}}{4}D'_g \\
\Psi &=& - {D_g+{4\over\sqrt{3}x}V\over 4+2x^2}~.
\eea
Again, regularity at $x=0$ requires $D_1=0$. Comparing with
Eqs.~(\ref{urad},\ref{Psirad}) gives $D_2=2A$.
In the  {\bf super-horizon regime}, $x\ll1$,  we obtain
\be
\Psi={A\over 3}, \quad D_g=-2A - {A\over 3\sqrt{3}} x^2, \quad
V={A\over 2\sqrt{3}} x~.
\ee
On {\bf sub-horizon scales}, $x\gg1$,  we find oscillating 
 solutions with constant amplitude and
with frequency of $k/\sqrt{3}$:
\bea
V&=&	{\sqrt{3}A\over 2} \sin(x) \\
D_g&=&	2A \cos(x)~,~~~ \Psi=-A\cos(x)/x^{2}~.
\eea
Note that also for radiation perturbations 
\[ D=-{A\over 3\sqrt{3}}x^2 \ll \Psi \]
is small on super horizon scales, $x\ll 1$. The perturbation amplitude is 
given by the largest gauge invariant perturbation variable. 
We conclude therefore that perturbations outside the Hubble
horizon are frozen to first order. Once they enter the horizon
they start to collapse, but pressure resists the gravitational
force and the radiation fluid fluctuations oscillate at constant
amplitude. The perturbations 
of the gravitational potential oscillate and decay like $1/a^2$
inside the horizon.

\subsection{ Adiabatic initial conditions}

Adiabaticity requires that the perturbations of all contributions to
the energy density are initially in thermal equilibrium. This fixes
the ratio of the density perturbations of different components. There is no
entropy flux and thus $\Ga=0$. Here we consider as a simple example
non relativistic matter and radiation perturbations.
Since the matter and radiation perturbations behave in the
same way on super-horizon scales,
\be
D_g^{(r)}=A+Bx^2, \quad
D_g^{(m)}=A'+B'x^2, \quad
V^{(r)} \propto V^{(m)} \propto x,
\ee
we may require a constant ratio between matter and radiation
perturbations. As we have seen in the previous section, 
inside the horizon ($x>1$) radiation
perturbations start to oscillate while matter perturbations keep
following a power law. On sub-horizon scales a constant ratio
can thus no longer be maintained. There are two interesting
possibilities: adiabatic and isocurvature perturbations. Here we
concentrate on adiabatic perturbations which seem to dominate 
the observed CMB anisotropies.

From $\Ga=0$ one easily
derives that two components with $p_i/\rho_i =w_i =$constant,
$i=1,2$, are adiabatically coupled if $(1+w_1)D_g^{(2)}=(1+w_2)D_g^{(1)}$. 
Energy conservation then  implies that also their
velocity fields agree, $V^{(1)}=V^{(2)}$. This result is also a
consequence of the Boltzmann equation in the strong coupling regime.
 We therefore require 
\be
V^{(r)}=V^{(m)} ,
\ee
so that the energy flux in the two fluids is coupled initially.

We restrict ourselves to a matter dominated backgrouns, the situation
relevant in the observed universe after equality. We first have to
determine the radiation perturbations during a {\em matter dominated
  era}. Since $\Psi$ is dominated by the matter contribution (it is
proportional to the background density of a given component), we 
have  $\Psi\simeq \mr{const}=\Psi_0$. We neglect the 
contribution from the sub-dominant radiation 
to $\Psi$. Energy--momentum conservation for radiation then
gives, with $x=k\eta$,
\bea
D_g^{(r)\prime} &=& -\frac{4}{3} V^{(r)}\\
V^{(r)\prime}    &=& 2\Psi + \frac{1}{4} D_g^{(r)} .
\eea

Now $\Psi$ is just a constant given by the matter
perturbations, and it acts like a constant source term.
The general solution of this system is then 
\bea
D_g^{(r)} &=& A \cos(c_sx)
	- \frac{4}{\sqrt{3}}B \sin(c_sx)
	+ 8 \Psi \l[\cos(c_sx)-1\r]\\
V^{(r)} &=& B \cos(c_sx)
	+ \frac{\sqrt{3}}{4}A \sin(c_sx)
	+ 2\sqrt{3} \Psi \sin(c_sx) ,
\eea
where $c_s=1/\sqrt{3}$ is the sound speed of radiation.
Our adiabatic initial conditions require
\be
\lim_{x\rightarrow 0} \frac{V^{(r)}}{x}=V_0
=\lim_{x\rightarrow 0} \frac{V^{(m)}}{x} < \infty .
\ee
Therefore $B=0$ and $V_0=A/4-2\Psi$. Using in addition $\Psi=3V_0$
(see (\ref{eq48})) we obtain
\bea
D_g^{(r)} &=& \frac{4}{3} \Psi \cos\l(\frac{x}{\sqrt{3}}\r)
	- 8 \Psi \label{Dad}\\
V^{(r)} &=& {1\over \sqrt{3}} \Psi \sin\l(\frac{x}{\sqrt{3}}\r)\\
D_g^{(m)} &=& -\Psi( 5 + \frac{1}{6} x^2) \\
V^{(m)} &=& {1\over 3} \Psi x .
\eea
On super-horizon scales, $x\ll 1$ we have 
\be 
	D_g^{(r)}  \simeq -\frac{20}{3} \Psi ~\mbox{ and }~~~ V^{(r)}
	\simeq \frac{1}{3}x \Psi~, \label{Dads}
\ee 
note that $D_g^{(r)} = (4/3)D^{(m)}_g$ and $V^{(r)}=V^{(m)}$ for adiabatic 
initial conditions.

Another possibility for the initial condition would be iso-curvature
initial conditions, where you have non-vanishing $D^{(r)}$, $D^{(m)}$
and $V^{(r)}$, $V^{(m)}$ which compensate each other in such a way
that $\Psi=0$ on super-horizon scales. The simplest inflationary
models do not lead to such perturbations and the observations imply
that they are not dominating the observed anisotropies in the CMB even
though they may contribute which could seriously hamper the 
determination of cosmological parameters with CMB anisotropies (see
e.g. \cite{Rob,Mood}

\section{Scalar field cosmology}
\label{scalarfield}
\index{scalar field}
We now consider the special case of a Friedmann universe filled with
self interacting scalar field matter. The action is given by
\be
 S = {1\over 16\pi G}\int d^4x\sqrt{|g|}R  +
     \int d^4x\sqrt{|g|}\left({1\over 2}\dd_\mu\vph\dd^\mu\vph - W(\vph)\right)
\ee
where $\vph$ denotes the scalar field and $W$ is the potential. The
energy momentum tensor is obtained by varying the action wrt the
metric $g_{\mu\nu}$, 
 \be
   T_{\mu\nu} = \dd_\mu\vph \dd_\nu\vph -
   \left[{1\over 2}\dd_\la\vph\dd^\la\vph + W \right]g_{\mu\nu}
\ee
The energy density $\rho$ and the energy flux $u$ are defined by
\be
 T^\mu_\nu u^\nu = -\rho u^\mu ~.
\ee
For the Friedmann background this gives
\be \rho = {1\over 2a^2}\dot\vph^2 + W  ~~~~~~~~ (u^\mu) = {1\over
a}( 1, \vec{0}) ~. \ee
The pressure is given by
\be T^i_j = p\de^i_j  \qquad  p = {1\over 2a^2}\dot\vph^2 - W  ~. \ee

We now define the scalar field perturbation,
\be  \label{defphi}
\vph = \bar{\vph} +\de\vph ~.  \ee
Clearly, the scalar field only generates scalar perturbations (to first order).
Inserting Eq.~(\ref{defphi}) in the definition of the energy velocity
perturbation $v$, 
\be (u^\mu) = {1\over a}( 1-A, -v,_i) \ee
and the energy density perturbation $\de\rho$,
\be \rho = \bar{\rho} +\de\rho ~, \ee
we obtain
\be \de\rho  ={1\over a^2}\dot{\bar{\vph}}\de\dot{\vph} - {1\over
a^2}\dot{\bar{\vph}}^2A + W,_\vph\de\vph  \ee
and
\be v = {k\over \dot{\bar{\vph}}}\left(\de\vph +\dot{\bar{\vph}}
k^{-1}B\right)~. \ee
From the stress tensor, $T_{ij} = \vph,_i  \vph,_j -
   \left[{1\over 2}\dd_\la\vph\dd^\la\vph + W \right]g_{ij}$ we find
\be p\pi_L =  {1\over a^2}\dot{\bar{\vph}}\de\dot{\vph} - {1\over
a^2}\dot{\bar{\vph}}^2A - W,_\vph\de\vph  \qquad  \mbox{and }\ \Pi = 0 ~.\ee

We now define the gauge invariant scalar field perturbation as the
value of $\de\vph$ in longitudinal gauge.
\be
\de\vph^{(gi)} = \de\vph^{(long)} = \de\vph + \dot{\bar{\vph}}\left(
         B-k^{-1}\dot H_T\right) = \de\vph - \dot{\bar{\vph}}\si~.
\ee
The last expression gives $\de\vph^{(gi)}$ in a generic gauge. It is clear
that this combination is gauge--invariant. This variable is very simply related
to the other gauge--invariant scalar variables. Short calculations give
\begin{eqnarray}
V &=&  k\de\vph^{(gi)}/ \dot{\bar{\vph}}\\
D_g &=& -(1+w)\left[4\Psi + 2{\dot a\over a}k^{-1}V- k^{-1}\dot V \right] \\
D_s &=& D_g+3(1+w)\Psi \\
\Ga &=& {2W,_\vph\over p\dot\rho}\left[\dot{\bar{\vph}}\rho D_s -
  \dot\rho \de\vph^{(gi)}\right]  \\  \Pi &=& 0~.
\end{eqnarray}
The last equation shows that the two Bardeen potentials are equal for
scalar field perturbations, $\Phi=\Psi$.
With this we can write the perturbed Einstein equations fully in terms
of the Bardeen potential $\Psi$ and $V$. Since we will need them
mainly to
discuss inflation where curvature plays a minor role, we write them
down here only for the case of vanishing spatial curvature. From
Eqs.~(\ref{scalcntr}) and (\ref{scaldyn}) one can easily generalize to
the case with curvature. 
\begin{eqnarray}
-3{\HH}\dot\Psi -(\dot{\HH}+ \HH^2-k^2)\Psi &=& 4\pi
G\dot\vph k^{-1}(\dot\vph V +a^2W,_\vph V) \\
\dot\Psi + {\HH}\Psi &=& 4\pi G\dot\vph^2k^{-1}V\\
\ddot\Psi + 3{\HH}\dot\Psi +(\dot{\HH}+2{\HH}^2)\Psi &=&
4\pi G\dot\vph k^{-1}(\dot\vph\dot V -\ddot\vph V -a^2W,_\vph V)
\end{eqnarray}
These lead to the following second order equation for the Bardeen
potential which we have discussed above:
\be  \ddot\Psi + 3({\HH}-\ddot\vph/\dot\vph)\dot\Psi
+(2\dot{\HH}-2{\HH}\ddot\vph/\dot\vph +k^2)\Psi =0
 \ee
or, using the definition $c_s^2= \dot p/\dot\rho$,
\be  \ddot\Psi + 3{\HH}(1+c_s^2)\dot\Psi +(2\dot{\HH}
+(1+3c_s^2){\HH}^2 +k^2)\Psi =0~.
 \ee
 As already mentioned above, this equation differs from the $\Psi$
 equation for a perfect fluid only in the last term proportional to
 $k^2$. This comes from the fact that the scalar field is not in a
 thermal state with fixed entropy, but it is in a fully coherent state
 ($\Ga\neq 0$) and field fluctuations propagate with the speed of
 light. On large scales, $k\eta\ll 1$ this difference is not relevant,
 but on sub--horizon scales it does play a certain role.

\section{Generation of perturbations during inflation}
\index{inflation}
\label{sec:inf}
 So far we have simply assumed some initial fluctuation amplitude $A$,
without investigating where it came from or what the $k$--dependence of $A$
might be. In this section we discuss the most common idea about
the generation of cosmological perturbations, namely their production
from the quantum vacuum fluctuations during an inflationary phase. The
treatment here is focused  mainly on getting the correct result with
as little effort as possible; we ignore several subtleties related,
\eg to the transition from quatum fluctuations of the field to
classical fluctuations in the energy momentum tensor. The idea is of
course that the sourse of metric fluctuations are the expectation
values of the energy momentum tensor operator of the scalar field. 

The basic idea is simple: A time dependent gravitational field
very generically leads to particle production, analogously to the
electron positron production in a classical, time dependent
electromagnetic field.

\subsection{Scalar perturbations}
The main result is the following: During inflation, the produced particles
 induce a perturbed gravitational field with a (nearly) scale invariant
 spectrum,
\be  \label{specin}
k^3|\Psi(k,\eta)|^2 = k^{n-1}\times\mr{const.} \quad \mbox{ with }
 \quad n\simeq 1~.
\ee

The quantity $k^3|\Psi(k,\eta)|^2$ is the squared amplitude of the
metric perturbation at comoving scale $\la=\pi/k$. To insure
that this quantity is small on a broad range of scales, so that
neither black holes are formed on small scales nor there are large
deviation from homogeneity and isotropy on large scales, we must
require $n\simeq 1$. These arguments have been put forward for the first
time by Harrison and Zel'dovich~\cite{HZ} (still before the advent of
inflation), leading to the name 
'Harrison-Zel'dovich spectrum' for a scale invariant perturbation
spectrum.

To derive the above result we consider a scalar field background dominated
by a potential, hence $a\propto |\eta|^{q}$ with $q\sim -1$.
Looking at the action of this system,
\[ S = \int dx^4\sqrt{|g|}\left(
   {R\over 16\pi G} +{1\over 2}(\nabla\vph)^2\right)~,
\]
it can be shown (see~\cite{Mukhanov:1992tc}) that the second order
perturbation of this action around the Friedmann solution is given by
\be \label{action}
 \de S =  \int dx^4\sqrt{|\overline g|}{1\over 2}(\dd_\mu v)^2
\ee
up to some total differential. Here $v$ is the perturbation variable
\be v = -{a\sqrt{\HH^2-\dot\HH}\over \sqrt{4\pi G}\HH}\ze \ee
introduced in Eq.~(\ref{defv}). Via the Einstein equations, this
variable can also be interpreted as representing the fluctuations in
the scalar field. Therefore, we quantize $v$ and assume that
initially, on small scales, $k|\eta|\ll 1$, $v$ is in the (Minkowski) quantum
vacuum state of a massless scalar field with mode function
\be \label{vin}
  v_\mr{in} = {v_0\over \sqrt{k}}\exp(ik\eta) ~.
\ee
The pre-factor $v_0$ is a $k$-independent constant which depends on
convention, but is of order unity.
From (\ref{cons}) we can derive
\[ (v/z)^{\textstyle\cdot} = {k^2 u\over z} ~,\]
where $z\propto a$ is defined in Eq.~(\ref{z}) and $u \propto
a\eta\Psi$ is given in Eq.~(\ref{defu}). On small scales, $k|\eta|\ll 1$,
this results in the initial condition for $u$
\be\label{uin} u_\mr{in} = {-iv_0\over k^{3/2}}\exp(ik\eta) ~.\ee

The evolution equation for $u$, (\ref{ueq2}), reduces in the case of
power law expansion, $a\propto |\eta|^q$ to
\be
\ddot u +( k^2 -{q(q+1)\over \eta^2})u = 0.
\ee
The solutions to this equation are of the form
$(k|\eta|)^{1/2}H^{(i)}_\mu(k\eta)$, where $\mu=q+1/2$ and
$H^{(i)}_\mu$ is the Hankel function of the $i$th kind ($i=1$ or $2$) of order
$\mu$. The initial condition (\ref{uin}) requires that only $H^{(2)}_\mu$
appears, so that we obtain
\[
 u = {\al\over k^{3/2}}(k|\eta|)^{1/2}H^{(2)}_\mu(k\eta) ~,
\]
where again $\al$ is a constant of order unity. We define the value of
the Hubble constant during inflation, which is nearly constant by
$H_i$. With $H=\HH/a \simeq 1/(|\eta| a)$ we then obtain  $a\sim
1/H_i|\eta|$. Eq.~(\ref{defu}) with the
Planck mass defiend by  $8\pi G=M^{-2}_P$ then gives 
\be
\Psi = {H_i\over 2M_{P}}u \simeq
  {H_i\over M_{P}} k^{-3/2}(k|\eta|)^{1/2}H^{(2)}_\mu(k\eta) ~.
\ee
 On small scales this is a simple oscillating function
while on large scales $k|\eta|\ll1$ it can be approximated by a
power law,
\be \Psi  \simeq  {H_i\over M_{P}}
k^{-3/2}(k|\eta|)^{1+q}  \simeq
  {H_i\over M_{P}} k^{-3/2} ~,~\mbox{ for }~ k|\eta|\ll 1~~~~.
\ee
Here we have used $\mu=1/2+q<0$ and $q\sim -1$.
This yields
\be\label{infspec}
k^3|\Psi|^2 \simeq  \left({H_i\over M_{P}}\right)^2~,
\ee
hence $n=1$. Detailed studies have shown that even though the
amplitude of $\Psi$ can still be severely affected by the transition
from inflation to the subsequent radiation era, the obtained spectrum
is very stable. Simple deviations from de Sitter inflation (like \eg
power law inflation), $q>-1$ lead to
slightly blue spectra, $n\gsim 1$.

\subsection{Vector perturbations}
In the simplest models of inflation where the only degrees of freedom
are the scalar field and the metric, no vector perturbations are generated.
But even if they are, subsequent evolution after inflation will lead
to their decay. In a perfect fluid background, $\Pi_{ij}=0$, vector
perturbations evolve according to Eq.(\ref{veccons}) which implies
\be
\Om \propto a^{3c_s^2-1} .
\ee
For a radiation fluid, $\dot{p}/\dot{\rho}=c_s^2\leq\nicefrac1/3$, this leads to a
non--growing vorticity. The dynamical Einstein equation~(\ref{vecdyn}) gives
\be
\si^{(V)}\propto a^{-2}~,
\ee
and the constraint (\ref{veccntr}) reads (at early times,
so that we can neglect curvature)
\be
\Om \sim (k\eta)^2\si^{(V)}.
\ee
Therefore, even if they are created in the very early universe on
super--horizon scales during an inflationary period, vector
perturbations of the metric decay and become soon entirely
negligible. Even if $\Om_i$ remains constant in a radiation
dominated universe, it has to be so small on relevant scales
at formation ($k\eta_{in}\ll1$) that we may safely neglect it.

Vector perturbations are irrelevant if perturbations have
been created at some early time, \eg during inflation. This result
changes completely when considering 'active perturbations' like for
example topological defects where vector perturbations contribute
significantly to the CMB anisotropies on large scales, see Ref.~\cite{report}.
Furthermore, it is interesting to note that vector perturbations do not
satisfy a wave equation and therefore will in no case show
oscillations. Vorticity simply decays with time.

\subsection{Tensor perturbations}

The situation is different for tensor perturbations. Again we
consider the perfect fluid case, $\Pi_{ij}^{(T)}=0$. Eq.
(\ref{tensdyn}) implies, if $\ka$ is negligible,
\be
\ddot H_{ij}+\frac{2\dot a}{a}\dot H_{ij}+k^2H_{ij}=0~. \label{tens}
\ee
 If the background has a power law evolution, $a\propto \eta^q$ this
 equation can be solved in terms of Bessel or Hankel functions. The
 less decaying  mode solution to Eq.~(\ref{tens}) is
$H_{ij} =e_{ij}x^{1/2-\b}J_{1/2-q}(x)$, where $J_\nu$ denotes the
Bessel function of order $\nu$, $x=k\eta$  and $e_{ij}$ is a transverse traceless
polarization tensor. This leads to
\bea
H_{ij}&=&\mr{const}\quad\mr{for}\;x\ll1\\
H_{ij}&=&\frac{1}{a}\quad\mr{for}\;x\gsim 1 ~.
\eea
One may also quantize the tensor fluctuations which represent
gravitons. Doing this, one obtains (up to log corrections a
scale invariant spectrum of tensor fluctuations from inflation:
For tensor perturbations the canonical variable is simple given by
$h_{ij} = M_PaH_{ij}$. 
The evolution equation for $h_{ij} =he_{ij}$ is of the form
\be \ddot h +(k^2+m^2(\eta)) h =0 ~,\label{tencan} \ee
where $m^2(\eta)=-\ddot a/a$. During inflation $m^2=-q(q-1)$ is
negative, leading to particle creation. Like for scalar perturbations,
the vacuum initial conditions are given on scales 
which are inside the horizon, $k^2\gg |m^2|$,
\[h_\mr{in} = {1\over\sqrt{k}}\exp(k\eta) ~~\mbox{ for }~ k|\eta|\gg  1.\]
Solving Eq.~(\ref{tencan}) with this initial condition, gives 
\[ h = {1\over \sqrt{k}}(k|\eta|)^{1/2}H^{(2)}_{q-1/2}(k\eta)~,\]
where $H^{(2)}_\nu$ is the Hankel function of degree $\nu$ of the second kind.
On super horizon scales, $H^{(2)}_{q-1/2}(k\eta)\propto (k|\eta|)^{q-1/2}$
this  results in $|h|^2 \simeq |\eta|(k|\eta|)^{2q-1}$. Using the
relation between $h_{ij}=he_{ij}$ and $H_{ij}$ one obtains the
spectrum of tensor perturbation generated during inflation. For exponential
inflation, $q\simeq -1$ one finds again a scale invariant spectrum for
$H_{ij}$ on super-horizon scales.
\be
 k^3|H_{ij}H^{ij}| \simeq (H_\mr{in}/M_P)^2 \propto k^{n_T} \quad
\mbox{ with } n_T \simeq 0~.
\ee

\section{Lightlike geodesics and CMB anisotropies}
\index{CMB anisotropies}
After decoupling, $\eta>\eta_\mr{dec}$, photons follow to a good approximation
light-like geodesics. The temperature shift of a Planck distribution
of photons is equal to the energy shift of any given photon, which is
independent of the photon energy (gravity is 'achromatic').

The unperturbed photon trajectory follows
\[
(x^\mu(\eta)) \equiv (\eta,\int_\eta^{\eta_0} \bn(\eta')d\eta'+\bx_0),
\]
where $\bx_0$ is the
photon position at time $\eta_0$ and $\bn$ is the (parallel transported)
photon direction. With respect to a geodesic basis $\l({\bf e}\r)_{i=1}^3$,
the components of $\bn$ are constant. If $\ka=0$ we may choose
${\bf e}_i=\dd/\dd x^i$; if $\ka\neq0$ these vector fields
are no longer parallel transported and therefore do not form a
geodesic basis ($\nabla_{{\bf e}_i}{\bf e}_j=0$).

Our metric is of the form
\bea
d\tilde{s}^2 &=& a^2 ds^2 \quad \mr{,with} \\
ds^2 &=& \l(\ga_{\mu\nu}+h_{\mu\nu}\r)dx^\mu dx^\nu ,
\quad \ga_{00}=-1, \, \ga_{i0}=0,\,\ga_{ij}=\ga_{ji}
\eea
as before.

We make use of the fact that light-like geodesics are conformally
invariant. More precisely, $ds^2$ and $d\tilde{s}^2$ have the same
light-like geodesics, only the corresponding affine parameters
are different. Let us denote the two
 affine parameters
by ${\la}$ and $\tilde\la$ respectively, and the tangent vectors to the
geodesic by
\be n = \frac{dx}{d\la}, \,\,\,\, \;  \tilde{n} = {1\over a}n =
\frac{dx}{d\tilde{\la}} \;\;, \;\;\; n^2 = \tilde{n}^2 = 0 \;, \;\;  n^0 =1
\;,\;\; {\bf n}^2 =1.
\ee
We set $n^0 = 1 +\de n^0$. The geodesic
equation for the perturbed metric
\be ds^2 =
(\ga_{\mu\nu}+h_{\mu\nu})dx^{\mu}dx^{\nu}  \ee
 yields, to first order,
\be
{d\over d\la}\de{n}^\mu = -\de\Ga^\mu_{\al\b}n^\al n^\b .
\ee
For the energy shift, we have to determine $\de n^0$. Since
$g^{0\mu}=-1\cdot \de_{0\mu}+\mr{first~order}$, we obtain
$\de\Ga^0_{\al\b}=-\nicefrac1/2 (h_{\al 0|\b}+h_{\b 0|\al}-\dot{h}_{\al\b})$,
so that
\be
{d\over d\la}\de{n}^0=h_{\al 0|\b}n^\b n^\al-
    \frac{1}{2}\dot{h}_{\al\b}n^\al n^\b .
\ee
Integrating this equation we use
  $h_{\al 0|\b} n^\b={d\over d\la} (h_{\al 0}n^\al)$,
so that the change of $n^0$ between some initial time $\eta_i$ and some
final time $\eta_f$ is given by
\be \de n^0 |_i^f = \left[h_{00} + h_{0j}n^j\right]_i^f -
   {1\over 2}\int_i^f\dot{h}_{\mu\nu}n^{\mu}n^{\nu}d\la  \; .
\label{2deltan}
\ee
On the other hand, the ratio of the energy of a photon measured by
some observer at $t_f$ to the energy emitted at $t_i$ is
\be
{E_f\over E_i} = \frac{(\tilde{n}\cdot u)_f}{(\tilde{n}\cdot u)_i}
    = {T_f\over T_i}
     \frac{(n\cdot u)_f}{(n\cdot u)_i}  \; , \label{Ef/Ei}
\ee
where $u_f$ and $u_i$ are the four-velocities of the observer and
emitter
respectively, and the factor $T_f/T_i$ is the usual (unperturbed)
redshift, which relates
 $n$ and $\tilde{n}$.
The velocity field of observer and emitter is given by
 \be u = (1-A)\dd_\eta +v^i\dd_i \; . \ee

An observer measuring
a temperature  $T_0$ receives photons that were emitted at the
time $\eta_{dec}$ of decoupling of matter and radiation, at the fixed
temperature $T_{dec}$. In first-order perturbation theory, we find the
following relation between the unperturbed temperatures $T_f$,
$T_i$,  the measurable temperatures $T_0=T_f+\de T_f$,
$T_{dec}=T_i+\de T_i$, and the photon
density perturbation:
\be {T_f \over T_i} =
    {T_0\over T_{dec}}\left(1 - {\de T_f\over T_f} + {\de T_i
\over T_i}\right) =
    {T_0\over T_{dec}}\left(1 - {1\over 4}\de^{(r)}|_i^f\right) \; ,
\ee
where $\de^{(r)}$ is the intrinsic density perturbation in the
radiation and we have used $\rho^{(r)}\propto T^4$ in the last
equality. Inserting the above equation
and Eq.~(\ref{2deltan}) into
Eq.~(\ref{Ef/Ei}),
and  using Eq.~(\ref{defpert}) for the definition of $h_{\mu\nu}$,
one finds, after  integration by parts \cite{Rfund} the following
result for scalar perturbations:
\be
 {E_f\over E_i} = {T_0\over T_{dec}}\left\{1-\left[ {1\over
4}D^{(r)}_g +
      V_j^{(b)}n^j  +\Psi+\Phi\right]_i^f +
    \int_i^f(\dot{\Psi}+ \dot{\Phi})d\la\right\}  \; .
\label{2deltaE}  \ee
Here $D_g^{(r)}$ denotes the density perturbation in the radiation
fluid, and  $V^{(b)}$ is the peculiar velocity of the baryonic matter
component (the emitter and observer of radiation).

Evaluating Eq.~(\ref{2deltaE}) at final time $\eta_0$ (today) and
initial time $\eta_{dec}$, we obtain the temperature difference
of photons coming from different directions $\bf n$ and ${\bf n}'$
 \be {\De T\over T} \equiv
{\De T({\bf n})\over T}- {\De T({\bf n}')\over T}
\equiv  {E_f\over E_i}({\bf n}) - {E_f\over E_i}({\bf n'}).
\ee
Direction independent contributions to  $E_f\over E_i$ do not
contribute to this difference. We also do not want to include the term
$V_j(\eta_0)n^j$ which simply describes the dipole due to our motion
with respect to the emission surface and which is much larger than the
contributions from the higher multipoles.  Therefore we can set
\be
{\Delta T({\bf n})\over T} =\left[ {1\over 4}D^{(r)}_g +
V_{j}^{(b)}n^j
+\Psi + \Phi\right](\eta_{dec},{\bf x}_{dec})
   + \int_{\eta_{dec}}^{\eta_0}(\dot{\Psi}+\dot{\Phi})(\eta,{\bf
    x}(\eta))d\eta~, \label{dT0}
\ee
where ${\bf x}(\eta)$ is the unperturbed photon position at time $\eta$ for
an observer at ${\bf x}_0$, and ${\bf x}_{dec}={\bf x}(\eta_{dec})$ (If
$\ka=0$ we simply have ${\bf x}(\eta)={\bf x}_0-(\eta_0-\eta){\bf n}$.).
The first term in Eq.~(\ref{dT0}) is the one we have discussed in the
previous chapter. It describes the intrinsic
inhomogeneities on the surface of  last scattering, due to acoustic
oscillations prior to decoupling. Depending on the initial conditions,
it can contribute significantly also on super-horizon scales.
This is especially important in the case of adiabatic initial
conditions. As we have seen in Eq.~(\ref{Dads}), in a dust $+$ radiation
universe with $\Om=1$, adiabatic initial conditions imply $D_g^{(r)}(k,\eta)
=-{20\over 3}\Psi(k,\eta)$ and $V^{(b)}=V^{(r)}\ll D_g^{(r)}$ for
$k\eta\ll 1$. With $\Phi=\Psi$ the the square bracket of
Eq.~(\ref{dT0}) therefore gives for adiabatic perturbations
\[
 \l({\Delta T({\bf n})\over T}\r)^{(OSW)}_{\rm adiabatic} =
    {1\over 3}\Psi(\eta_{dec},{\bf x}_{dec})
\]
on super-horizon scales. The contribution to ${\delta T\over T}$ from
the last scattering surface on very large scales is called the
'ordinary Sachs--Wolfe effect' (OSW). It has been derived for the first
time by Sachs and Wolfe~\cite{SW} in 1967.
For isocurvature perturbations,  the
initial condition $D_g^{(r)}(k,\eta)\ra 0$ for ${\eta\ra 0}$ is satisfied
and  the contribution of $D_g$ to the ordinary Sachs--Wolfe effect
can be neglected.
\[
 \l({\Delta T({\bf n})\over T}\r)^{(OSW)}_{\rm isocurvature}
    = 2\Psi(\eta_{dec},{\bf x}_{dec})~.
\]
The second term in (\ref{dT0}) describes
the relative motion of emitter and  observer. This is the Doppler
contribution to the  CMB anisotropies. It appears on the same
angular scales as the acoustic term; we  call the sum of
the acoustic and Doppler contributions ``acoustic peaks''.

The last two terms are due to the inhomogeneities in the spacetime
geometry; the first contribution determines the change in the photon
energy due to the difference of the gravitational potential at the
position of emitter and observer. Together with the part contained in
$D_g^{(r)}$ they represent the ``ordinary'' Sachs-Wolfe  effect. The
integral accounts for red-shift or blue-shift caused by the
time dependence of the gravitational field along the  path of the
photon, and represents the so-called integrated Sachs-Wolfe (ISW)
effect. In a $\Om=1$, pure dust universe, the Bardeen potentials are
constant and there is no integrated Sachs-Wolfe effect; the blue-shift
which the photons acquire by falling into a gravitational potential is
exactly canceled by the redshift induced by climbing out of it. This
is no longer true in a universe with substantial radiation
contribution, curvature or a cosmological constant.

The sum of the ordinary Sachs--Wolfe term and the integral is the full
Sachs-Wolfe contribution (SW).

For {\bf vector} perturbations $\de^{(r)}$ and $A$ vanish and Eq.~(\ref{Ef/Ei})
leads to
\be (E_f/E_i)^{(V)} = (a_i/a_f)[1 - V_j^{(m)}n^j|_i^f +
  \int_i^f\dot{\si}_jn^jd\la]   ~. \label{2dev} \ee
We obtain a Doppler term and a gravitational contribution.
For {\bf tensor} perturbations, i.e. gravitational waves,  only the
gravitational part remains:
\be (E_f/E_i)^{(T)} = (a_i/a_f)[1 -   \int_i^f\dot{H}_{lj}n^ln^jd\la]
   ~. \label{2det} \ee
Equations (\ref{2deltaE}), (\ref{2dev}) and (\ref{2det}) are the
manifestly gauge invariant results for the energy shift of photons due
to scalar, vector and tensor perturbations. Disregarding again the dipole
contribution due to our proper motion, Eqs.~(\ref{2dev},\ref{2det})
 imply the vector and tensor temperature fluctuations
\bea
\left({\Delta T({\bf n})\over T}\right)^{(V)} &=&
  V_j^{(m)}(\eta_{dec},{\bf x}_{dec})n^j +
  \int_i^f\dot{\si}_j(\eta,{\bf x}(\eta))n^jd\la  \label{dTV}\\
\left({\Delta T({\bf n})\over T}\right)^{(T)} &=&
-   \int_i^f\dot{H}_{lj}(\eta,{\bf x}(\eta))n^ln^jd\la \label{dTT}
   ~.  \eea
Note that for models where initial fluctuations have been laid down in
the very early universe, vector perturbations are irrelevant as we
have already pointed out. In this sense Eq.~(\ref{dTV}) is here
mainly for completeness. However, in models where perturbations are
sourced by some inherently inhomogeneous component ({\em e.g.} topological
defects, see Ref.~\cite{report}) vector perturbation can be important.

\section{Power spectra}
\index{power spectra}

One of the basic tools to compare models of large scale structure
having stochastic initial fluctuations
with observations are power spectra. They are the ``harmonic
transforms'' of the two point correlation functions. If the perturbations
of the model under consideration are Gaussian (a relatively generic
prediction from inflationary models), then the power spectra
contain the full statistical information of the model.

Let us first consider the power spectrum of dark matter,
\be
P_D(k)=\l\lan\l| D_g^{(m)}\l(\bk,\eta_0\r)\r|^2\r\ran ~.
\ee
Here $\lan~\ran$ indicates a statistical average, ensemble average,
 over ``initial
conditions'' in a given model. $P_D(k)$ is usually compared with
the observed power spectrum of the galaxy distribution. This is
clearly problematic since it is by no means evident what the ratio of
these two spectra should be. This problem is known under the name of
'biasing' and it is very often simply assumed that the dark matter and
galaxy power spectra differ only by a constant factor. The hope is
also that on sufficently large scales, since the  evolution of both, galaxies 
and dark matter is governed by gravity, their power spectra should not
differ too much. This hope seems to be reasonably justified~\cite{SDSS}.

The power spectrum of velocity perturbations satifies the relation
\be
P_V(k)=\l\lan\l| {\bf V}\l(\bk,\eta_0\r)\r|^2\r\ran
\simeq H_0^2 \Om^{1.2}P_D(k)k^{-2}~.
\ee
For $\simeq$ we have used that 
$|kV|(\eta_0) = \dot{D}_g^{(m)}(\eta_0) \sim H_0\Om^{0.6}D_g$
on sub-horizon scales (see \eg~\cite{peebles}).

The spectrum we can be both, measured and calculated to the best
accuracy is the CMB anisotropy power spectrum. It is defined as follows:
$\De T/T$ is a function of position $\bx_0$, time $\eta_0$ and
photon direction $\bn$. We develop the $\bn$--dependence in
terms of spherical harmonics. We will suppress the argument $\eta_0$
and often also $\bx_0$ in the
following calculations. All results are for today ($\eta_0$) and here
($\bx_0$). By statistical homogeneity statistical averages over an
ensemble of realisations (expectation values) are supposed
to be independent of position. Furthermore, we assume that the process 
generating the initial perturbations is statistically isotropic. Then,
the off-diagonal correlators of the expansion coefficients $a_{\ell m}$
vanish and we have
\be
\frac{\De T}{T}\l(\bx_0,\bn,\eta_0\r)
=\sum_{\ell,m} a_{\ell m}(\bx_0) Y_{\ell m}(\bn), \quad
\l\lan a_{\ell m}\cdot a_{\ell'm'}^*\r\ran 
= \de_{\ell\ell'}\de_{mm'}C_\ell
\ee
The $C_\ell$'s are the CMB power spectrum.

The two point correlation function is related to the $C_\ell$'s by
\bea
\l\lan \frac{\De T}{T}(\bn)\frac{\De T}{T}(\bn')\r\ran_{\bn\cdot\bn'=\mu}
=  \sum_{\ell,\ell',m,m'}\l\lan a_{\ell m}\cdot a_{\ell'm'}^*\r\ran
	 Y_{\ell m}(\bn) Y_{\ell' m'}^*(\bn') =  && \nonumber \\
\sum_\ell C_\ell \underbrace{\sum_{m=-\ell}^\ell 
   Y_{\ell m}(\bn) Y_{\ell m}^*(\bn')}_{\frac{2\ell+1}{4\pi}
   P_\ell(\bn\cdot\bn')}
= \frac{1}{4\pi}\sum_\ell (2\ell+1)C_\ell P_\ell(\mu) ,\qquad && \label{correl}
\eea
where we have used the addition theorem of spherical harmonics
for the last equality;  the $P_\ell$'s are the Legendre polynomials.

Clearly the $a_{lm}$'s from scalar, vector and tensor perturbations
are uncorrelated,
\be
\l\lan a_{\ell m}^{(S)} a_{\ell'm'}^{(V)} \r\ran
=\l\lan a_{\ell m}^{(S)} a_{\ell'm'}^{(T)} \r\ran
=\l\lan a_{\ell m}^{(V)} a_{\ell'm'}^{(T)} \r\ran
=0 .
\ee

Since vector perturbations decay, their contributions, the $C_\ell^{(V)}$,
are negligible in models where initial perturbations have been
laid down very early, \eg, after an inflationary period. Tensor
perturbations are constant on super-horizon scales and perform damped
oscillations once they enter the horizon.

Let us first discuss in somewhat more detail scalar perturbations.
We specialize to the case $\ka=0$ for simplicity.
We suppose the initial perturbations to be given by a spectrum,
\be
\l\lan\l|\Psi\r|^2\r\ran k^3=A^2 k^{n-1}\eta_0^{n-1} . \label{inspec}
\ee
We multiply by the constant $\eta_0^{n-1}$, the actual comoving size of the 
horizon, in order to keep $A$
dimensionless for all values of $n$. $A$ then represents the amplitude of 
metric perturbations at horizon scale today, $k=1/\eta_0$.

On {\em super-horizon scales} we have, for {\em adiabatic} perturbations:
\be
\frac{1}{4}D_g^{(r)} = -\frac{5}{3} \Psi +\OO((k\eta)^2), \quad
V^{(b)}=V^{(r)}=\OO(k\eta)
\ee

The dominant contribution on super-horizon scales (neglecting the 
integrated Sachs--Wolfe effect $\int \dot{\Phi}-\dot{\Psi}$~) is then
\be
\frac{\De T}{T}(\bx_0,\bn,\eta_0) = \frac{1}{3} \Psi(x_\mr{dec}, 
\eta_\mr{dec}).		\label{sw}
\ee

The Fourier transform of (\ref{sw}) gives
\be
\frac{\De T}{T}(\bk,\bn,\eta_0)  = \frac{1}{3} \Psi(k, \eta_\mr{dec}) \cdot
	e^{i\bk\bn\l(\eta_0-\eta_\mr{dec}\r)}~.
\ee

Using the decomposition
\[
e^{i\bk\bn\l(\eta_0-\eta_\mr{dec}\r)} =
   {\sum_{\ell=0}^\infty (2\ell+1)i^\ell
   j_\ell(k(\eta_0-\eta_\mr{dec})) P_\ell( \widehat{\bk}\cd\bn)}~,
\]
where $j_\ell$ are the spherical 
Bessel functions, we obtain
\bea
\lefteqn{\l\lan \frac{\De T}{T}(\bx_0,\bn,\eta_0)
 \frac{\De T}{T}(\bx_0,\bn',\eta_0) \r\ran}\\
 &=& \frac{1}{V} \int d^3x_0 \l\lan\frac{\De T}{T}(\bx_0,\bn,\eta_0)
 \frac{\De T}{T}(\bx_0,\bn',\eta_0) \r\ran \nonumber \\
&=&\frac{1}{(2\pi)^3}\int d^3k \l\lan\frac{\De T}{T}(\bk,\bn,\eta_0)
 \l(\frac{\De T}{T}\r)^*(\bk,\bn',\eta_0) \r\ran  \nonumber\\
& =& \frac{1}{(2\pi)^3 9}\int d^3k \l\lan\l|\Psi\r|^2\r\ran 
	\sum_{\ell,\ell'=0}^\infty
 (2\ell+1)(2\ell'+1) i^{\ell-\ell'} \nonumber \\ &&  \cdot
 j_\ell(k(\eta_0-\eta_\mr{dec}))
 j_{\ell'}(k(\eta_0-\eta_\mr{dec}))P_\ell(\hat{\bk}\cd\bn) \cdot
P_{\ell'}(\hat{\bk}\cd\bn')~.
\eea
In the second equal sign we have used the unitarity of the Fourier
transformation. Inserting $P_\ell(\hat{\bk}\bn) =
 \frac{4\pi}{2\ell+1} \sum_m Y_{\ell m}^*(\hat{\bk})Y_{\ell m}(\bn)$
and  \\   $P_{\ell'}(\hat{\bk}\bn')=
 \frac{4\pi}{2\ell'+1}\sum_{m'} Y_{\ell' m'}^*(\hat{\bk})Y_{\ell' m'}(\bn')$,
integration over the directions $d\Om_{\hat{k}}$ 
 gives $\de_{\ell\ell'}\de_{mm'}\sum_m Y_{\ell m}^*(\bn)Y_{\ell m}(\bn')$.\\
Using as well $\sum_m Y_{\ell m}^*(\bn)Y_{\ell m}(\bn')=\frac{2\ell+1}{4\pi}
P_\ell(\mu)$, where  $\mu=\bn\cdot\bn'$, we find
\bea
\lefteqn{\l\lan \frac{\De T}{T}(\bx_0,\bn,\eta_0)
 \frac{\De T}{T}(\bx_0,\bn',\eta_0) \r\ran_{\bn\bn'=\mu}
 =} \nonumber \\ &&
\sum_\ell \frac{2\ell+1}{4\pi} P_\ell(\mu) \frac{2}{\pi}
\int\frac{dk}{k} \l\lan\frac{1}{9}|\Psi|^2\r\ran k^3
j_\ell^2(k(\eta_0-\eta_\mr{dec})) .
\eea

Comparing this equation with~Eq.~(\ref{correl})
we obtain for {\em adiabatic perturbations}
on scales $2\le \ell$ $\ll$ 
$\chi(\eta_0-\eta_\mr{dec})/\eta_\mr{dec}$ $\sim 100$
\be
C_\ell^{(SW)} \simeq C_\ell^{(OSW)} \simeq 
 \frac{2}{\pi} \int_0^\infty \frac{dk}{k}
 \l\lan\l|\frac{1}{3}\Psi\r|^2
 \r\ran k^3 j_\ell^2 \l(k\l(\eta_0-\eta_\mr{dec}\r)\r).
\label{clsw}
\ee

If $\Psi$ is a pure power law as in Eq.~(\ref{inspec}) and we set
$k(\eta_0-\eta_\mr{dec})\sim k\eta_0$,  the integral (\ref{clsw}) can
be performed analytically. For the ansatz (\ref{inspec}) one finds 
\be
C_\ell^{(SW)} = \frac{A^2}{9} \frac{\Ga(3-n)\Ga(\ell-\frac{1}{2}+\frac{n}{2})}{
2^{3-n}\Ga^2(2-\frac{n}{2})\Ga(\ell+\frac{5}{2}-\frac{n}{2})}
\quad\mbox{ for }  -3<n<3~ . \label{clswsol}
\ee

Of special interest is the {\em scale invariant} or
Harrison--Zel'dovich spectrum, $n=1$ (see Section~\ref{sec:inf}).  It leads to

\be
\ell(\ell+1)C_\ell^{(SW)} = \mr{const.} \simeq
\l\lan\l(\frac{\De T}{T}(\vth_\ell)\r)^2\r\ran~,~~~~ 
	\vth_\ell\equiv \pi/\ell~.
\ee
This  is precisely (within the accuracy of the experiment) the
behavior  observed by the DMR experiment aboard the satellite
COBE \cite{DMR}.

Inflationary models predict very generically a HZ spectrum (up to
small corrections). The DMR discovery has therefore been
regarded as a great success, if not a proof, of inflation.
There are other models like topological defects \cite{PST,ZD,Aetal}
or certain string
cosmology models \cite{dgsv} which also predict scale--invariant,
\ie, Harrison Zel'dovich spectra of fluctuations. These models do
however not belong to the class investigated here, since in these
models perturbations are induced by seeds which evolve non--linearly
in time.

For isocurvature perturbations, the main contribution on large
scales comes from the integrated Sachs--Wolfe effect and (\ref{clsw})
is replaced by
\be
C_\ell^{(ISW)} \simeq \frac{8}{\pi} \int \frac{dk}{k}
k^3 \l\lan \l| \int_{\eta_\mr{dec}}^{\eta_0}  \dot{\Psi}(k,\eta)
j_\ell^2(k(\eta_0-\eta))d\eta\r|^2\r\ran . \label{clisw}
\ee
Inside the horizon $\Psi$ is roughly constant (matter dominated).
Using the ansatz (\ref{inspec}) for $\Psi$ inside the horizon
and setting the integral in (\ref{clisw})
$\sim2\Psi(k,\eta=1/k) j_\ell^2(k\eta_0)$, we obtain again
(\ref{clswsol}), but with $A^2/9$ replaced by $4A^2$. For a fixed
amplitude $A$ of perturbations, the Sachs--Wolfe
temperature anisotropies coming from isocurvature perturbations are
therefore about a factor of $6$ times larger than those coming from
adiabatic perturbations.

On smaller scales, $\ell\gsim 100$ the contribution to $\De T/T$
is usually dominated by acoustic oscillations, the first two terms in
Eq.~(\ref{dT0}).  Instead of 
(\ref{clisw}) we then obtain
\bea
\lefteqn{C_\ell^{(AC)}\simeq} \nonumber \\
&& \frac{2}{\pi} \int_0^\infty \frac{dk}{k}
k^3\l\lan\l|\frac{1}{4}D^{(r)}(k,\eta_\mr{dec})j_\ell(k\eta_0)
+V^{(r)}(k,\eta_\dec)j_\ell'(k\eta_0)\r|^2\r\ran~.
\eea

To remove the SW contribution from $D_g^{(r)}$ we have simply replaced
it by $D^{(r)}$ which is much smaller than $\Psi$ on super-horizon
scales and therefore does not contribute to the SW terms.
On sub-horizon scales $D^{(r)}\simeq D_g^{(r)}$ and $V^{(r)}$ are oscillating
like sine or cosine waves depending on the initial conditions.
Correspondingly the $C_\ell^{(AC)}$ will show peaks and
minima. On very small scales they are damped by the photon
diffusion which takes place during the recombination process (see
contribution by A. Challinor).

\begin{figure}[ht]
\centering
\includegraphics[width=8cm]{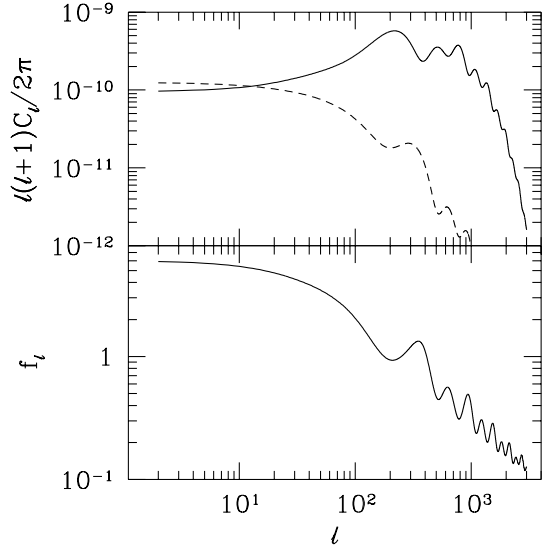}
\caption{A COBE normalized sample adiabatic (solid line) and isocurvature 
(dashed line) CMB anisotropy spectrum,  $\ell(\ell+1)C_\ell$, are shown on the 
top panel. The quantity
shown in the bottom panel is the ratio of temperature fluctuations for fixed 
value of $A$ (from Kanazawa et al.~\protect\cite{Kanaza}).}\label{adisofig}
\end{figure}

For gravitational waves (tensor fluctuations), a formula analogous to
(\ref{clswsol}) can be derived,
\be
C_\ell^{(T)}=\frac{2}{\pi}\int dk k^2 \l\lan\l|\int_{\eta_\dec}^{\eta_0}
d\eta \dot{H}(\eta,k) \frac{j_\ell(k(\eta_0-\eta))}{(k(\eta_0-\eta))^2}
\r|^2\r\ran\frac{(\ell+2)!}{(\ell-2)!} .
\ee

\begin{figure}[ht]
\centering
\includegraphics[width=8cm,angle=-90]{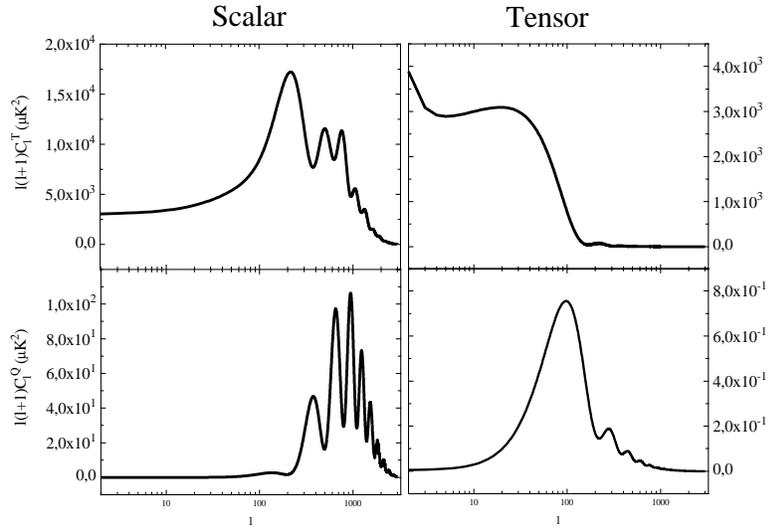}
\caption{\label{STfig}Adiabatic scalar and tensor 
 CMB anisotropy spectra are shown (top panels).
The bottom panels show the corresponding polarization spectra (see
course by A. Challinor). 
(from \protect\cite{melvit}).}
\end{figure}

To a very crude approximation we may assume $\dot{H}=0$ on super-horizon
scales and $\int d\eta \dot{H}j_\ell(k(\eta_0-\eta)) 
	\sim H(\eta=1/k)j_\ell(k\eta_0)$. For a pure power
law,
\be
k^3\l\lan\l|H(k,\eta=1/k)\r|^2\r\ran=A_T^2 k^{n_T}\eta_0^{-n_T} ,
\ee
one obtains
\bea
C_\ell^{(T)} &\simeq & \frac{2}{\pi} \frac{(\ell+2)!}{(\ell-2)!} A_T^2
\int \frac{dx}{x} x^{n_T} \frac{j_\ell^2(x)}{x^4} \nonumber \\
&=& \frac{(\ell+2)!}{(\ell-2)!} A_T^2 
	\frac{\Ga(6-n_T)\Ga(\ell-2+\frac{n_T}{2})}{
    2^{6-n_T}\Ga^2(\frac{7}{2}-n_T)\Ga(\ell+4-\frac{n_T}{2})} .
\eea
For a scale invariant spectrum ($n_T=0$) this results in
\be
\ell(\ell+1)C_\ell^{(T)} \simeq \frac{\ell(\ell+1)}{(\ell+3)(\ell-2)}
	A_T^2{8\over 15\pi}~ .
\ee
The singularity at $\ell=2$ in this crude approximation is not real,
but there is some enhancement of $\ell(\ell+1)C_\ell^{(T)}$
at $\ell\sim2$ see Fig.~\ref{STfig}).

Since tensor perturbations decay on sub-horizon scales, $\ell \gsim 60$,
they are not very sensitive to cosmological parameters.

Again, inflationary models (and topological defects) predict a scale
invariant spectrum of tensor fluctuations ($n_T \sim 0$).

On very small angular scales, $\ell\gsim 800$, fluctuations are damped
by collisional damping (Silk damping). This effect has to be discussed
with the Boltzmann equation for photons which is presented in detail
in the course by A. Challinor.

\section{Some remarks on perturbation theory in braneworlds}
\index{braneworlds}
Since there has been so much interest in them recently, let me finally
make some remarks on perturbation theory of five dimensional
braneworlds. I shall just present some relatively simple aspects
without derivation. A thorough discussion of braneworlds is given in
the course by R. Maartens. Different aspects of the perturbation
theory of braneworlds can be found in the growing literature on the
subject~\cite{bpert,ours}. 

The bulk background metric of a five dimensional braneworld has three
dimensional spatial slices which homogeneous and isotropic, hence spaces of
constant curvature. Its line element is therefore of the form

\be \label{bbackgd}
ds^2 = -n^2(t,y)dt^2 +a^2(t,y)\ga_{ij}dx^idx^j + b^2(t,y)dy^2 =g_{AB}dx^Adx^B
\ee
Perturbations of such a spacetime can be decomposed into scalar, vector
and tensor modes with respect to the three dimensional spatial slices
of constant curvature. One can always choose the so-called {\bf generalized
longitudinal gauge} such that the perturbations of the metric are
given as follows:

\bea 
ds^2 &=& -n^2(1+2\Psi)dt^2 +a^2\l[(1-2\Phi)\ga_{ij}+2H_{ij}\r]dx^idx^j 
+ b^2(1+2C)dy^2  \nonumber \\ &&
-2nbBdtdy -2na\Si_idx^idt +ab2\EE_idx^idy   ~.  \label{bpert}
\eea
Here $\Si_i$ and $\EE_i$ are divergence free vector fields, vector
perturbations  and $H_{ij}$ is a divergence free traceless symmetric
tensor, the tensor perturbation. $\Psi,\Phi,C$ and $B$ are four
scalar perturbations.

It can be shown that this choice determines the gauge
completely. One can actually define gauge invariant perturbation
variables which reduce to the ones above in the generalized
longitudinal gauge~\cite{ours}. Writing down the perturbed Einstein
equations for these variables in the most general case is quite
involved. These equations can be found in Ref.~\cite{ours}, but I
don't want to repeat them here. I just discuss their general
structure in the case of an empty bulk. 
It is clear that $\Psi$ and $\Phi$ correspond to the Bardeen
potentials of four dimensional cosmology, $\Si_i$ is the four
dimensional vector perturbation and $H_{ij}$ represents four
dimensional gravitational waves. $C$ and $B$ as well as $\EE_i$ are new
degrees of freedom which are not present in the four dimensional theory.

If we assume vanishing
perturbations of the bulk energy momentum tensor e.g. if the bulk is
anti de Sitter like in the Randall Sundrum model~\cite{RSII} called
  RSII in what follows (see course by R.~Maartens),
the perturbation equations reduce to
\bea
\Box_5(\Psi+\Phi) &=&0 \label{ws}\\
\Box_5\Si_i &=& 0 \label{wv}\\
\Box_5H_{ij} &=& 0   \label{wt}
\eea
and all the other perturbation variables are determined by constraint
equations. Here $\Box_5$ is the five dimensional d'Alembertian with
respect to the background metric~(\ref{bbackgd}). This structure of
the equations is to be expected: Gravitational waves in $d$ spacetime
dimensions are a spin 2 field with respect to the group of rotations
$SO(d-2)$ since they are 
massless (see e.g.~\cite{Wein}). For $d=5$, $d-2=3$ they therefore have 5
degrees of freedom. These correspond exactly to the one scalar (\ref{ws}), two
vector  (\ref{wv}) and two tensor  (\ref{wt}) degrees of freedom with
respect to the 3-dimensional slices of conrant curvature. These free
massless degrees of freedom obey the wave equations above. 

The perturbed Israel junction conditions (see course by R.~Maartens)
then determine boundary conditions for the behavior of the
perturbations at the brane position(s).

As an example, I write down the vector and tensor perturbation
equations and their bulk solutions for the RSII model which has
only one brane. There the bulk is a five dimensional anti-de
Sitter spacetime, and we can choose coordinates so that the background
metric has the form
\be \label{rs2}
ds^2 =\l({L\over y}\r)^2\l[-dt^2 +\de_{ij}dx^idx^j+dy^2 \r]
\ee 
and
\be \label{rsdAle}
\Box_5 = -\dd_t^2+\lap +\dd_y^2 -{3\over y}\dd_y~,
\ee
where $\lap$ denotes the three dimensional spatial Laplacian.
For an arbitrary mode which satisfies this wave equation we make the ansatz 
\[
\phi(t,\bx,y) = \exp(i(\bk\cd\bx-\om t))\phi(\om,\bk,y)~.
\]
We then obtain a Bessel differential equation for $\phi(\om,\bk,y)$ with
general solution
\bea
\phi &=& A(\om,\bk)(ym)^2J_2(my) + B(\om,\bk)(ym)^2Y_2(my)~,  \quad
  m^2 =\om^2-k^2~, \nonumber \\
 &=& \phi_A + \phi_B ~.
\eea 
These modes are normalizable in the sense that 
\[ \int_{y_b}^\infty |\phi|^2\sqrt{-g}dy < \infty ~.\]
Here $y_b>0$ is the brane position.
The Israel junction condition for the vector and tensor modes for RSII
become
\bea
-\dd_yH_{ij}(y=y_b) &=& \ka_5^2\Pi_{ij}(y_b) \\
 -\Si_i(y_b) &=& \ka_5^2\Pi_i(y_b)~, \label{sol5}
\eea
where $\Pi_{ij}$ respectively $\Pi_i$
are the tensor rsp. vector contribution to the anisotropic stresses on
the brane. If the latter vanish, the junction condition simple requires
\[ B(\om,\bk) =  A(\om,\bk)J_2(my_b)/Y_2(my_b) ~.\]
This result has been derived for the tensor mode by Randall and
Sundrum~\cite{RSII}. For scalar perturbations the situation is
somewhat more complicated since there is an additional degree of
freedom which is the perturbed position of the brane, $y_b \ra
y_b+\ep$. It is  rather subtle to take this brane bending correctly
into account.  A very interesting work showing that this effect is actually
most relevant to obtain the correct Newtonian limit in the RSII model can
be found in Ref.~\cite{GT}

Let us also briefly discuss the zero--mode. From the brane point of
view, the modes discussed here represent waves (particles) which
couple only to the energy momentum tensor of the brane and which obey
a dispersion relation $\om^2-k^2=m^2$, hence the parameter $m$ of the
solutions~(\ref{sol5}) is their mass. 
In the limit $m\ra 0$ the solutions turn into power laws in $y$,
\be \phi_0(y) = Ay^4 + B ~.\ee
Of these modes, for tensor and vector perturbations, the $B$ mode is
normalizable. For scalar perturbations the situation is more
complicated. For a mode to be 'normalizable' we want all the
perturbations to be normalizable,
\[\int_{y_b}^\infty |\phi|^2\sqrt{-g}dy < \infty ~, \quad \mbox{ for }
\phi = \Phi~, ~\Psi ~,~ 
C  \mbox{ and } B ~. \]
But from the constraints one obtains for $m=0$ $C\propto y^2$ which
is not normalizable. This mode diverges logharitmically  and
only converges due to the oscillations of the Bessel functions if
$m\neq 0$.

If all the five graviton zero--modes are normalizable, like
in all compact braneworlds, e.g. in models with two branes, the vector
and scalar mode lead to three additional degrees of freedom (to the
usual two four-dimensional graviton modes) which couple via the
junction conditions to the brane energy momentum tensor and spoil the
phenomenology of the model. The resulting  four 
dimensional  gravity is not Einstein  gravity but Brans--Dicke or even
more complicated. As an example, in Ref.~\cite{DK} the contribution
from the scalar zero--mode~(\ref{ws}) to the 
gravity wave emission from a binary pulsar is calculated and shown to
be in contradiction with observations. This problem is well-known from
Kaluza Klein theories, it is the so called moduli problem. 
The way out of it is usually to render the modes  massive.
There are different suggestions how this can be achieved for the
scalar mode. One possibility is the so called Goldberger--Wise
mechanism~\cite{GW} which shows that under certain circumstances a
bulk scalar field can do the job. But certainly, a physically
acceptable braneworld is defined only together with its mechanism how
to get rid of such unwanted modes (see also Ref.~\cite{LS}. 

The advantage of the RSII model is that the scalar gravity wave zero--mode
is not normalizable in this model and therefore it does not
contribute. This very promising property of the RSII model has let to its
popularity. It seems to induce the correct four dimensional Einstein
gravity on the brane, in the cosmological context this property is still
maintained at sufficiently low energies.

\section{Conclusions}
In this course I have given an introduction to cosmological
perturbation theory. Perturbation theory is an important tool
especially to calculate CMB anisotropies and polarisation since these
are very small and can be determined reliably within linear cosmological
perturbation theory. To determine the evolution of the cosmic matter
density, linear perturbation theory has to be complemented with the
theory of weakly non-linear Newtonian gravity and with N-body
simulations. To finally understand the formation of galaxies 
non-gravitational highly non-linear physics, like heating and cooling
mechanisms, dissipation, nuclear reactions etc. have to be taken into
account. This very difficult subject is still in its infancy. 

To make progress in our understanding of braneworlds, linear
perturbation theory can also be most helpful. We can use it to
determine e.g. the propagating modes of the gravitational field on the
brane, light deflection and redshift in weak gravitational fields 
and the Newtonian limit. The condition that linear perturbations on
the brane at low energy and large distances reduce to those resulting
from Einstein gravity is non--trivial and has, to my knowledge, not
yet been fully explored to limit  braneworld models.

\acknowledgement
I thank the organizers for a well structured school in a most
beautiful environment.


\printindex
\end{document}